\documentclass[sigconf]{acmart}
\usepackage{popets}
\usepackage[disable]{todonotes}
\usepackage{adjustbox}
\usepackage{algorithm}
\usepackage{enumitem}
\usepackage{array} 
\usepackage{booktabs}
\usepackage{graphicx}
\usepackage{tikz}
\usetikzlibrary{positioning,fit,backgrounds,calc,arrows.meta}
\usepackage{microtype}
\usepackage{xurl}
\definecolor{capReq}{HTML}{1976D2}
\definecolor{capRep}{HTML}{228833}
\definecolor{capAck}{HTML}{DD6622}
\usepackage{pifont}
\newcommand{\cmark}{\ding{51}}
\newcommand{\xmark}{\ding{55}}

\usepackage{float}
\usepackage{dblfloatfix}
\usepackage{placeins}
\usepackage{caption}
\usepackage{amsthm}
\usepackage{algpseudocode}

\setcopyright{popets}
\copyrightyear{2026}

\acmYear{2026}
\acmVolume{YYYY}
\acmNumber{X}
\acmDOI{}
\acmISBN{}
\acmConference{Proceedings on Privacy Enhancing Technologies}
\begin{document}

% The "title" command has an optional parameter, allowing the author to define a
% "short title" to be used in page headers.
\title[...]{Hidden Services Protocol for Mixnets}

% The "author" command and its associated commands are used to define the
% authors and their affiliations.
\author{Nicolas Constantinides}
\authornote{Corresponding author.}
\affiliation{%
  \institution{Unaffiliated}
  \country{}}
\email{}

\author{Mahdi Rahimi}
\affiliation{%
  \institution{COSIC, KU Leuven}
  \city{Leuven}
  \country{Belgium}}
\email{}

\author{Stavros Nonis}
\affiliation{%
  \institution{Unaffiliated}
  \country{}}
\email{}

\renewcommand{\shortauthors}{Constantinides et al.}

\begin{abstract}
%Even when message contents are encrypted, the Internet Protocol reveals communication metadata: who talks to whom, when, and how often.
%Mix networks (mixnets) hide these relationships by sending each packet through a sequence of independently chosen relays that delay and cryptographically transform it before forwarding, so observers cannot easily link a packet's entry to its exit.
%In systems such as Nym, every user attaches persistently to one \emph{entry gateway}, which is their long-term point of presence in the network.
%Learning that gateway is therefore almost as valuable as learning the user: deanonymization no longer requires breaking the whole mix path and instead reduces to attacking that single attachment point.
%Nym also lacks a hidden-service mechanism in which both a client and a service can communicate while keeping their network locations hidden.

Mix networks (mixnets) provide network-level privacy by routing each communication packet through a sequence of intermediaries, called mixnodes, that randomly delay and cryptographically transform packets before forwarding them, making it difficult for observers to link mixnet entries to exits. While this mechanism protects sender privacy from both external adversaries and the receiver, existing mixnets lack a secure and practical protocol that simultaneously protects receiver (destination) privacy, particularly from the sender. We close this gap.

We introduce the first practical and secure hidden-service protocol for mixnets, providing receiver privacy alongside sender anonymity. Our design builds on Single-Use Reply Blocks (SURBs), which enable anonymous replies without revealing the receiver's address. We show, however, that existing approaches to using SURBs expose two practical attacks that can compromise sender or receiver anonymity when the opposing party controls only a single mixnode. We develop defenses against both vulnerabilities.
Building on these defenses, we introduce \emph{NymHS}, a secure and practical hidden-service protocol for mixnets that supports anonymous service discovery, authenticated bidirectional sessions, and asynchronous SURB replenishment. We implement NymHS on the open-source Nym codebase and evaluate its practicality through web-browsing experiments across 118 websites, measuring each of the 27 configurations three times ($9{,}558$ page loads). Our results demonstrate that hidden services can be deployed efficiently over mixnets. In particular, increasing the Sphinx payload from $2\,\mathrm{KiB}$ to $10\,\mathrm{KiB}$ reduces mean page-load latency by approximately $5.2\times$ and decreases communication overhead from $21.7\%$ to $4.3\%$ relative to the current Nym baseline.

\end{abstract}

% Keywords. The author(s) should pick words that accurately describe the work
% being presented. Separate the keywords with commas.
%, Sphinx, Single-Use Reply Blocks (SURBs), gateway discovery vulnerability}
\maketitle

\section{Introduction}
%The Internet Protocol exposes communication metadata---who communicates with whom, when, and how often---even when payloads are encrypted~\cite{diaz2021nym}.
%Anonymous communication networks hide these relationships~\cite{syverson2004tor,diaz2021nym,danezis2003mixminion,zantout2011i2p}.
%Among them, mix networks (mixnets)~\cite{chaum1981untraceable,piotrowska2017loopix,diaz2021nym,diaz2004taxonomy,kesdogan1998stop} offer some of the strongest guarantees even in the presence of a passive global adversary that observes every network link: each packet traverses an independently chosen sequence of mix nodes that delay and cryptographically transform it before forwarding thereby making challenging correlating input and output traffic~\cite{rahimi2025parsan}.

The Internet Protocol inherently exposes communication metadata---who communicates with whom, when, and how often---even when payloads are encrypted, as encryption protects only the content of communications~\cite{diaz2021nym}. Anonymous communication networks aim to protect this metadata~\cite{chaum1981untraceable,syverson2004tor,diaz2021nym,danezis2003mixminion,zantout2011i2p}. Among these systems, mix networks (mixnets)~\cite{chaum1981untraceable,piotrowska2017loopix,diaz2021nym,diaz2004taxonomy,kwon2020xrd,van2015vuvuzela,mahdi2026OptiMix,mahdi2025lamp,mahdi2024larmix} provide some of the strongest privacy guarantees, even against a passive global adversary that observes every network link. This protection arises because each packet traverses an independently selected sequence of intermediaries, called mixnodes, which randomly delay and cryptographically transform packets before forwarding them, making input--output traffic correlation challenging~\cite{rahimi2025parsan}.

%This work focuses on Loopix~\cite{piotrowska2017loopix}, a low-latency Stop-and-Go mixnet built on the Sphinx packet format~\cite{Danezis2009Sphinx}.
%Sphinx provides constant-size, layered-encrypted packets whose headers carry hop-by-hop forwarding instructions and whose payloads carry application data.
%To resist traffic analysis, Loopix emits packets on two independent streams: a \emph{loop cover traffic stream} of dummy packets that return to their sender, and a \emph{real traffic stream} that sends queued application packets when available and otherwise sends further loop cover.
%Because both streams are scheduled independently of whether application data are present, the observable transmission pattern does not reveal when a user has real messages to send (Section~\ref{sec:background}).

In this work, we focus on Loopix-like mixnets~\cite{piotrowska2017loopix}, in which mixnodes are partitioned into layers and each mixnode forwards packets only to mixnodes in subsequent layers. Each mixnode provides mixing by delaying received packets for a random interval sampled according to the stop-and-go mixing mechanism~\cite{kesdogan1998stop}. Loopix builds on the Sphinx packet format~\cite{Danezis2009Sphinx}, which provides constant-size, layered-encrypted packets whose headers contain hop-by-hop forwarding instructions and whose payloads carry application data.
To resist traffic analysis, Loopix generates packets through two independent streams: a \emph{loop cover traffic stream}, consisting of dummy packets sent by Loopix users that eventually return to them, and a \emph{real traffic stream}, which sends queued application packets when available and otherwise transmits additional loop cover traffic at the same rate. Because both streams are scheduled independently of whether application data are available, the observable transmission pattern does not reveal when a user has real messages to send, thereby providing unobservability.

Nym~\cite{diaz2021nym} operationalizes the Loopix design using a five-stage topology consisting of entry gateways, three mix layers, and exit gateways. The three mix layers perform the mixing functionality of Loopix, while the entry and exit gateways act as interfaces between users and the mixnet. In particular, each user persistently connects to a single entry gateway, which serves as their long-term entry point into the mixnet.
Beyond one-way anonymous messaging from a sender to a receiver, Nym supports Single-Use Reply Blocks (SURBs). A SURB allows a party that wishes to receive replies---the \emph{SURB creator}---to provide another party---the \emph{replier}---with a disposable return path, enabling the replier to respond without learning the creator's identity or the gateways and mixnodes traversed by the return path. The intended anonymity is therefore one-sided: the creator's location remains hidden from the replier, while the reply must still enter the mixnet through an entry gateway.

To construct this return path, the creator precomputes a unique routing token, represented by a Sphinx reply header, together with per-hop encryption secrets for each mixnode on the path and a separate key for protecting the application payload. The creator then provides this information to the replier. To send a reply, the replier first encrypts the application content using the reply key, then prepends a replier-generated acknowledgment SURB together with a small amount of metadata. The packet is subsequently processed using the per-hop secrets such that, as it traverses the return path, each mixnode removes the corresponding encryption layer. By the time the packet reaches the creator's gateway, the outer layers have been removed. The gateway can then extract the acknowledgment SURB and reinject it into the mixnet to support reliable delivery, while the application content remains encrypted under the reply key and can be decrypted only by the creator.

%As deployed in Nym, however, that package is enough to break anonymity for both sides.
%Because the routing ticket is unique and known to the creator in advance, a malicious creator can place a mix they control as the first mix hop of the return path and simply watch for their own ticket.
%Seeing it reveals which gateway injected the reply---the replier's long-term entry point.
%The same first-mix-hop trick applies to acknowledgments: the replier chooses the acknowledgment SURB's path, so a malicious replier can plant the first mix hop of that path and recognize its own ticket when the creator's gateway sends the acknowledgment, exposing the creator's long-term entry gateway.
%A dual failure sits at the other end of the path.
%Because Nym ships \emph{all} of the per-hop secrets to the replier, the replier can also compute, in advance, the exact payload ciphertext that will appear at every hop---including the last mix hop before the creator's gateway (in Nym, the third mix layer).
%A malicious replier that also controls that mix can therefore recognize ``this is that reply'' by matching those predicted bytes (or the corresponding secret the hop derives locally) and learn the next hop, which is the creator's long-term entry gateway (Section~\ref{sec:first-hop-surb-attack}).
%In both cases, gateway discovery is decisive: once the adversary knows the victim's long-term entry gateway, deanonymizing that user no longer requires breaking the mix path and instead reduces to compromising or flooding that gateway.

Although SURBs are deployed in the Nym mixnet, their current design introduces anonymity risks for both communicating parties. First, because the routing token associated with a SURB is unique and known to the SURB creator in advance, a malicious creator can construct the return path so that a mixnode under their control serves as its first mix hop and then monitor that mixnode for the corresponding token. Observing the token reveals which gateway injected the reply, thereby exposing the replier's long-term entry gateway.

Conversely, Nym provides the replier with all per-hop payload-key seeds associated with the SURB. A malicious replier that also controls the mixnode serving as the final hop of the return path can therefore identify the corresponding reply by deriving the payload transformation expected at that hop and matching it against the observed packet. This reveals the creator's long-term entry gateway (Section~\ref{sec:first-hop-surb-attack}). In both cases, discovering the victim's gateway is decisive: once the adversary learns the victim's long-term entry gateway, deanonymization no longer requires compromising the entire mix path, but instead reduces to compromising or flooding that gateway.

Beyond the vulnerabilities in the current deployment of SURBs, a second limitation is the absence of hidden services for mixnets. Currently, contacting a Nym application typically requires either prior knowledge of its network address or the use of a long-lived nickname tied to a fixed entry gateway, allowing that gateway to associate the nickname with the underlying user (Section~\ref{relatedwork}). A hidden service, by contrast, keeps both the client and the service location hidden while eliminating the need to know the receiver's exact network address in advance, a property that is particularly important for applications such as anonymous feedback and whistleblowing~\cite{Mazieres1998Nymserver}.

We address these vulnerabilities and limitations. \textbf{First}, to enable secure SURB-based communication, we introduce a new protocol that does not expose mix-derived payload-key seeds to the replier. Instead, the prior-hop seeds are embedded within the layered-encrypted final routing block, where only the creator's gateway can recover them. The gateway can then reverse the cryptographic transformations applied by the mixnodes, extract the acknowledgment SURB, and inject it back into the mixnet, while preventing the final mix hop from recognizing the reply using disclosed secrets or predicted ciphertext.

\textbf{Second}, we introduce \emph{replier-anonymous SURBs}, which conceal the creator's precomputed routing token inside an outer Sphinx packet and reveal it only at a designated swap hop. Consequently, a malicious creator monitoring the first mix hop of the inner route observes traffic arriving from the swap path rather than directly from the replier's gateway.
Building on these primitives, we \textbf{third} design and implement \emph{NymHS}, the first hidden-service protocol for the Nym mixnet, providing authenticated publication, anonymous discovery, pseudonymous bidirectional sessions, and asynchronous SURB replenishment, with both control and application traffic routed entirely through the mixnet.

\textbf{Fourth}, we extend the open-source Nym codebase to implement our hidden-service protocol and evaluate NymHS through static web browsing across 118 websites under 27 configurations---nine Sphinx payload sizes by three rewrap depths---each repeated three times, for $9{,}558$ page loads. Our performance evaluation shows that increasing the Sphinx payload from $2\,\mathrm{KiB}$ to $10\,\mathrm{KiB}$ reduces mean page-load latency by approximately $5.2\times$ and payload overhead from $21.7\%$ to $4.3\%$, while increasing the rewrap depth adds only $0.3$--$0.6$\,s of latency (Section~\ref{sec:evaluation}).

%Our contributions are as follows.
%\begin{itemize}[nosep]
%\item We analyze Nym SURB replies and acknowledgment-based reliability (Section~\ref{sec:sphinx-surb}).
%\item We identify first-mix-hop and last-mix-hop attacks that expose a replier's or creator's gateway while controlling a single mix node, and show their impact on Nym applications (Sections~\ref{sec:first-hop-surb-attack} and~\ref{relatedwork}).
%\item We withhold mix-derived payload-key seeds from the replier, embedding them for gateway recovery and preventing last-mix-hop matching (Section~\ref{sec:last-hop-fix}).
%\item We introduce Sphinx-compatible replier-anonymous SURBs that prevent first-mix-hop header matching (Section~\ref{sec:replier-anon}).
%\item We design and implement NymHS, the first mixnet hidden-service protocol built on these defenses (Section~\ref{sec:hidden-service}).
%\item We evaluate NymHS under varying payload sizes and rewrap depths, showing that larger payloads cut fragmentation cost while replier anonymity adds modest latency (Section~\ref{sec:evaluation}).
%\end{itemize}

%Section~\ref{sec:background} reviews Loopix, Nym, and SURBs; Section~\ref{sec:first-hop-surb-attack} presents the attacks; Sections~\ref{sec:requirements}--\ref{methodology} develop NymHS and its security properties; Section~\ref{sec:evaluation} reports experiments; and Section~\ref{relatedwork} discusses related work.

In particular, our contributions are as follows.
\begin{itemize}[nosep]
\item We analyze the security of Nym SURB replies and their use for acknowledgment-based reliability (Section~\ref{sec:sphinx-surb}).
\item We identify first-mix-hop and last-mix-hop attacks that expose the replier's or creator's gateway when the adversary controls only a single mixnode, and demonstrate their impact on Nym applications (Sections~\ref{sec:first-hop-surb-attack} and~\ref{relatedwork}).
\item We prevent last-mix-hop matching by withholding mix-derived payload-key seeds from the replier and embedding them for recovery by the creator's gateway (Section~\ref{sec:last-hop-fix}).
\item We introduce Sphinx-compatible replier-anonymous SURBs that prevent first-mix-hop header matching (Section~\ref{sec:replier-anon}).
\item We design and implement NymHS, the first hidden-service protocol for mixnets built on these defenses (Section~\ref{sec:hidden-service}).
\item We evaluate NymHS across varying payload sizes and rewrap depths, showing that larger payloads substantially reduce fragmentation overhead while replier anonymity incurs only modest additional latency (Section~\ref{sec:evaluation}).
\end{itemize}

\noindent\textbf{Roadmap.}
Section~\ref{sec:background} reviews Loopix, Nym, and SURBs; Section~\ref{sec:first-hop-surb-attack} presents the identified attacks; Sections~\ref{sec:requirements}--\ref{methodology} develop NymHS and analyze its security properties; Section~\ref{sec:evaluation} reports the experimental evaluation; and Section~\ref{relatedwork} discusses related work.

\color{black}{}
\section{Background}
\label{sec:background}

\color{black}
\color{black}
%This section reviews the Nym mixnet architecture and the Sphinx SURB construction.
%Appendix~\ref{app:sphinx} gives the formal header and payload notation used throughout the paper, including Nym-specific differences from Danezis and Goldberg~\cite{Danezis2009Sphinx}.

This section reviews the Nym mixnet architecture, the Sphinx packet format, and its SURB construction. For a more detailed representation of the Sphinx packet format, please refer to Appendix~\ref{app:sphinx}.

\subsection{Nym Mixnet Architecture}
\label{sec:loopix}

Nym is a Loopix-like mixnet organized into five layers: an entry-gateway layer, three mix layers, and an exit-gateway layer. The gateway layers serve as interfaces between users and the mixnet without performing mixing, while the three intermediate layers constitute the core mixnet and perform the transformations required to anonymize traffic. In particular, entry gateways authenticate users, verify whether they are authorized to access the system, and, if so, inject their traffic into the mixnet; importantly, they do not perform Sphinx packet processing. Sphinx processing is instead performed by the remaining four layers. Each of the three mix layers removes one Sphinx layer, recovers the next forwarding hop, and randomly delays the packet to provide mixing. The exit gateway removes the final Sphinx layer, delivers the message\footnote{We use the terms packet and message interchangeably throughout this paper.} to its destination, sends acknowledgments back to the sender through the mixnet, and may buffer traffic for offline users.

An important design feature of Nym is that each user persistently connects to a single entry gateway, which serves as that user's long-term point of presence in the network. The gateway also temporarily stores packets destined for offline clients and forwards them when the clients reconnect. Discovering this gateway is therefore particularly consequential for deanonymization: once a user's long-term attachment point is known, the adversary can target that gateway directly rather than having to infer the complete route traversed by the user's messages to associate the client with its communication.

In Nym, messages are transmitted using fixed-size Sphinx packets~\cite{Danezis2009Sphinx}, each of which is routed independently through the mix layers. Packets belonging to the same communication may therefore traverse different routes, making end-to-end traffic correlation more difficult for a network adversary. At each mix layer, the corresponding mixnode removes one cryptographic layer, learns only the next forwarding hop, and applies a sender-selected delay drawn from an exponential distribution with mean $\mu=50\,\mathrm{ms}$~\cite{oldenburg2024mixmatch}.\footnote{Detailed Sphinx header and payload formats are provided in Appendix~\ref{app:sphinx}.}

Following Loopix, Nym senders maintain two independent traffic streams~\cite{piotrowska2017loopix}. On the \emph{real traffic stream}, transmission opportunities occur according to a Poisson process with a mean inter-departure time of $20\,\mathrm{ms}$. At each opportunity, the sender transmits the head-of-line application packet if the queue is non-empty and otherwise emits a loop cover packet. A separate \emph{loop cover traffic stream} generates additional dummy packets that eventually return to their sender, with a mean inter-departure time of $200\,\mathrm{ms}$ at both clients and mixnodes~\cite{oldenburg2024mixmatch}. Because both streams operate independently of whether application data are queued, a passive observer cannot readily infer when a user is transmitting real traffic.

\subsection{Sphinx and SURB-Based Replies}
\label{sec:sphinx-surb}

Nym relies on the Sphinx packet format~\cite{Danezis2009Sphinx}, which provides constant-size packets and layered anonymous routing with packet indistinguishability, replay protection, and bitwise unlinkability between hops. A Sphinx packet consists of a header and a payload. The header received at hop~$i$ is $M_i=(\alpha_i,\beta_i,\gamma_i)$, where $\alpha_i$ is an ephemeral Diffie--Hellman group element, $\beta_i$ is an encrypted routing block, and $\gamma_i$ is a header-integrity tag; the accompanying payload ciphertext is denoted by~$\delta_i$. To maintain a fixed packet size, Sphinx pads headers and payloads when necessary. We denote by $\mathrm{Pad}_\pi$ the Sphinx padding operation to a fixed payload length and by $\pi_k$ the keyed payload permutation under key~$k$. Sphinx additionally relies on several key-derivation functions, a routing-block stream cipher, an integrity MAC, a blinding map, and size parameters $(\eta,r,\kappa)$, which we detail in Appendix~\ref{app:sphinx-notation}; the complete Sphinx header construction is provided in Appendix~\ref{app:sphinx-header}.

\subsubsection*{Sphinx Processing}

We model $h_\tau$, $h_\mu$, $h_\rho$, $h_b$, and $h_\sigma$ as random-oracle hash functions. Under this model, upon receiving $(M_i,\delta_i)$, mixnode~$n_i$ along the packet's route uses its private key $x_{n_i}$ to compute the Diffie--Hellman shared secret $s_i=\alpha_i^{x_{n_i}}$, verifies $\gamma_i\stackrel{?}{=}\mu(h_\mu(s_i),\beta_i)$, and aborts if verification fails. To prevent replay attacks, it also derives the replay tag $\tau_i=h_\tau(s_i)$ and drops the packet if $\tau_i$ has already been observed; otherwise, it records~$\tau_i$, preventing the same header from being processed twice.

The mixnode then derives the routing-mask stream $\rho_i=\rho(h_\rho(s_i))$, the payload-key seed $\sigma_i=h_\sigma(s_i)$, and the payload key $k_i^\pi=h_{\mathsf{k}}(\sigma_i)$. Extending $\beta_i$ by one routing-block step and removing the corresponding mask reveals the next-hop information:
\[
\beta_i'=\beta_i\Vert 0^\eta,
\qquad
m_i\Vert n_{i+1}\Vert\gamma_{i+1}\Vert\beta_{i+1}= \beta_i'\oplus\rho_i,
\]
where $m_i$ contains per-hop routing metadata, including the forwarding delay and a flag, $|m_i\Vert n_{i+1}\Vert\gamma_{i+1}|=\eta$, and $|\beta_{i+1}|=r\eta$. The mixnode then reblinds the group element as $\alpha_{i+1}=\alpha_i^{h_b(\alpha_i,s_i)}$, removes one payload layer as $\delta_{i+1}=\pi^{-1}_{k_i^\pi}(\delta_i)$, and forwards $(\alpha_{i+1},\beta_{i+1},\gamma_{i+1},\delta_{i+1})$ to~$n_{i+1}$. The final hop performs the corresponding processing without introducing an additional delay, recovers the inner payload and its destination, and forwards the message accordingly.

SURBs extend Sphinx to support anonymous replies~\cite{Danezis2009Sphinx}. A SURB is defined as
\[
\mathcal{S}=\bigl(n_0,,M_0,,\sigma_0,\ldots,\sigma_{\nu-1},,\tilde{k}\bigr),
\]
where $M_0$ is a precomputed Sphinx header for the first mix hop~$n_0$, $\sigma_0,\ldots,\sigma_{\nu-1}$ are the payload-key seeds associated with a reply route of length~$\nu$, and $\tilde{k}\in\{0,1\}^\kappa$ is an independent reply-encryption key rather than a seed. Upon receiving the SURB, the replier derives each $k_i^\pi$ from the supplied seeds as described above and uses $\tilde{k}$ exclusively to encrypt the application-layer reply.\footnote{The requester supplies SURBs before any reply is transmitted; applications that exchange SURBs are discussed in Section~\ref{relatedwork}.} Each reply fragment consumes one SURB: the replier constructs $\delta_0$ using the supplied seeds and $\tilde{k}$, injects $(M_0,\delta_0)$ at~$n_0$, and can send the reply without learning the recipient's identity or complete return route.

\subsubsection*{SURB Transmission}
\label{sec:surb-transmission}

Because SURBs are single-use, a fragmented reply consisting of $n$ pieces requires a pool $\{\mathcal{S}_1,\ldots,\mathcal{S}_n\}$ with
\[
\mathcal{S}_i=\bigl(n_0^{(i)},\,M_0^{(i)},\,\sigma_0^{(i)},\ldots,\sigma_{\nu-1}^{(i)},\,\tilde{k}_i\bigr).
\]
The replier's complete message is divided into fragments $\{f_1,\ldots,f_n\}$; each fragment $f_i$ carries a marshaled fragment header $\mathsf{FH}_i$ containing a set identifier $\mathit{sid}$, the fragment count~$n$, the fragment index~$i$, and an optional successor-set link when a message spans more than $255$ fragments.

For each $i$, the replier constructs a delivery-confirmation packet $\mathrm{SURB\text{-}ACK}_i$ over a fresh acknowledgment route and forms
\[
p_i
=
\mathrm{SURB\text{-}ACK}_i
\Vert
H(\tilde{k}_i)
\Vert
\mathrm{Enc}^{\mathsf{AES}}_{\tilde{k}_i}(\mathsf{FH}_i\Vert f_i),
\]
where $H(\tilde{k}_i)$ is a key digest that allows the creator's client to later locate~$\tilde{k}_i$ and decrypt~$\mathsf{FH}_i\Vert f_i$. The replier pads $p_i'=\mathrm{Pad}_\pi(p_i,\ell)$ to the Sphinx payload size~$\ell$ and applies the payload transformations
\[
\delta_0^{(i)}
=
\pi_{k_0^{\pi,(i)}}
\circ\cdots\circ
\pi_{k_{\nu-1}^{\pi,(i)}}
(p_i')
\]
under the keys derived from $\sigma_0^{(i)},\ldots,\sigma_{\nu-1}^{(i)}$. It then sends $(M_0^{(i)},\delta_0^{(i)})$ to~$n_0^{(i)}$. The recipient gateway forwards $\mathrm{SURB\text{-}ACK}_i$ into the mixnet and delivers the remaining ciphertext to the client; missing acknowledgments trigger retransmission using a fresh SURB. Once online, the client matches $H(\tilde{k}_i)$ against its stored keys, recovers~$\tilde{k}_i$, and decrypts the fragment.

\paragraph{Design rationale.}
In the original Sphinx SURB~\cite{Danezis2009Sphinx}, the replier receives the precomputed header together with the single reply key~$\tilde{k}$; mix-derived payload keys are not disclosed. Nym instead includes every seed $\sigma_0,\ldots,\sigma_{\nu-1}$ so that, when the reply reaches the creator's gateway, the mix-derived payload layers can be reversed while the application ciphertext protected under~$\tilde{k}$ remains sealed. The gateway can then extract the embedded $\mathrm{SURB\text{-}ACK}$ and send it back to the replier, confirming that the packet was received even when the receiver is offline and the $\tilde{k}$-protected fragment must be buffered until the client reconnects. This separation therefore enables reliable communication with offline receivers: the gateway can return the $\mathrm{SURB\text{-}ACK}$ while buffering the sealed fragment for subsequent delivery. However, disclosing the mix-derived seeds to the replier also introduces the vulnerabilities discussed in Section~\ref{sec:first-hop-surb-attack}.

\section[Problem and SURB Path Attacks]{Problem Statement and Malicious SURB Path Attacks}
\label{sec:first-hop-surb-attack}

Standard SURBs enable anonymous replies, but they do not protect endpoint anonymity when either the SURB creator or the replier is malicious and additionally controls a single mixnode. To characterize this limitation, we first define our threat model and then present two attacks---the first-mix-hop and last-mix-hop attacks---that expose a user's gateway. We subsequently discuss the consequences of gateway discovery and formulate the design problem that motivates our defenses.

\paragraph{Threat model.}
We adopt Nym's infrastructure-aware threat model~\cite{diaz2021nym}. In particular, our design considers (1)~a computationally bounded global passive adversary (GPA) capable of observing traffic across all network links, (2)~a mixnode adversary that operates malicious mixnodes, (3)~an adversary that controls a subset of the gateways, and (4)~a malicious user that additionally controls a single mixnode.\footnote{We do not otherwise consider active attacks that are already detectable or mitigated by mechanisms provided by Nym.}

\paragraph{Malicious first-mix-hop attack.}
The first attack combines a malicious SURB creator with a mixnode under the creator's control, which the creator deliberately selects as the first mix hop~$n_0$ of the reply path. Because the adversary operates this mixnode, it can inspect packets arriving at the hop and recognize the reply associated with its own SURB. Operating the mixnode is essential for this attack: inter-node links are protected using Noise (Section~\ref{sec:loopix}), preventing a passive observer from reading Sphinx headers. The malicious creator, however, already knows the unique header $M_0=(\alpha_0,\beta_0,\gamma_0)$ associated with the SURB and can therefore match $\alpha_0$, $\beta_0$, or $\gamma_0$ against packets observed at the controlled mixnode. A successful match immediately reveals the gateway directly upstream of~$n_0$, thereby exposing the replier's long-term entry gateway.

The same attack applies to acknowledgments in the opposite direction. A malicious client acting as the replier constructs a $\mathrm{SURB\text{-}ACK}$ whose first mix hop is a mixnode under its control. By recognizing the corresponding header when the acknowledgment reaches that mixnode, the adversary identifies the gateway from which it originated and thereby exposes the original SURB creator's gateway.

\paragraph{Malicious last-mix-hop attack.}
A dual attack combines a malicious replier with a malicious mixnode positioned at the last mix hop before the creator's gateway, corresponding to the third mix layer in Nym. Because a Nym SURB provides the replier with every mix-derived seed $\sigma_i$ (Section~\ref{sec:sphinx-surb}), the colluding replier and mixnode can recognize the corresponding reply in two ways.

First, from the supplied seeds, the replier derives all $k_i^\pi=h_{\mathsf{k}}(\sigma_i)$ and can therefore predict the $\approx2\mathrm{KiB}$ payload ciphertext that will appear at the controlled hop. The malicious mixnode can then directly match the observed payload against this predicted ciphertext. Second, the malicious mixnode derives the payload-key seed $\sigma_i=h_\sigma(s_i)$ from its Diffie--Hellman shared secret with the packet and compares it against the seeds supplied in the SURB. Either method uniquely identifies the reply and reveals its next hop, namely the creator's gateway.

Unlike in the first-mix-hop attack, the malicious replier cannot choose which mixnode in the last layer is selected, as Nym selects this node uniformly from roughly $40$ candidates. Nevertheless, application messages commonly consume many independently routed SURBs, one for each fragment. Consequently, even an adversary controlling only one last-layer mixnode obtains repeated opportunities to observe and recognize a reply, making the attack practical over repeated communication.

\paragraph{Gateway compromise.}
Once either attack reveals a user's gateway, an adversary that subsequently compromises that gateway can deanonymize the attached client. A compromised gateway observes every delivered fragment and can associate the client's IP address with the corresponding binding tag $H(\tilde{k})$. Because the replier already knows $H(\tilde{k})$, it can match this value against the gateway's records and identify the intended recipient. Such compromise need not rely on exploiting the gateway's software: bribery~\cite{karakostas2024blockchain}, coercion, or legal compulsion may provide equivalent access. Thus, an adversary with comparatively limited capabilities---a malicious user, control of a single mixnode, and subsequent access to the identified gateway---can ultimately deanonymize a communicating endpoint.

\paragraph{Active flooding attack.}
Gateway discovery also enables Nym's active flooding attack. An adversary possessing a pool of SURBs can inject a large number of replies toward the victim's gateway within a short time window, thereby amplifying traffic patterns and facilitating correlation. As documented by Nym, however, this attack requires prior knowledge of the target gateway.\footnote{\url{https://nym.com/docs/network/mixnet-mode/anonymous-replies}} NymHS does not attempt to prevent flooding itself; instead, it prevents the gateway discovery required to mount the attack against an otherwise anonymous endpoint.

\paragraph{Consequences for applications.}
Any protocol that exchanges unmodified Nym SURBs with potentially untrusted parties (Section~\ref{relatedwork}) inherits these vulnerabilities. Once an endpoint's gateway is exposed, its anonymity effectively reduces to the security of that single gateway, undermining the protection provided by the mix path. This work therefore addresses two complementary objectives: (i)~strengthening SURBs against malicious first- and last-mix-hop attacks while preserving compatibility with Sphinx, and (ii)~building a hidden-service protocol for Nym that enables mutually anonymous communication without revealing either endpoint's gateway.

\section{Design Requirements}
\label{sec:requirements}

To construct a hidden-service protocol that securely resists both malicious first- and last-mix-hop attacks, we define the following design requirements.

\begin{description}

\item[\textbf{R1 (Replier anonymity).}]
Reply packets transmitted using a SURB must not reveal the replier's network attachment point, i.e., the replier's gateway in Nym. In particular, a malicious SURB creator controlling the first mix hop of either the reply path or a $\mathrm{SURB\text{-}ACK}$ path must not be able to recognize the corresponding packets as they enter the mixnet.

\item[\textbf{R2 (Anonymous service discovery).}]
Clients must be able to discover and contact a hidden service using only its public pseudonym. The discovery protocol must not reveal the association between the service's pseudonym and its network attachment point to any party, while preserving the anonymity of the client.

\item[\textbf{R3 (Mutual endpoint anonymity).}]
Neither the client nor the hidden service should reveal to any other party the association between its public pseudonym and its network attachment point (e.g., IP address or entry gateway).

\item[\textbf{R4 (Authenticated bidirectional communication).}]
Mutually distrusting parties must be able to authenticate exchanged messages and maintain bidirectional communication using pseudonymous identifiers rather than network identifiers.

\item[\textbf{R5 (Long-lived sessions).}]
Interactive applications require asynchronous bidirectional communication over extended periods without relying on persistent return paths. Unlike circuit-based anonymous networks that reuse fixed paths, NymHS should construct a fresh anonymous route for every transmitted packet while still supporting long-lived sessions.

\item[\textbf{R6 (Sphinx compatibility).}]
The strengthened reply mechanism should preserve the Sphinx packet format and routing semantics. Mixnodes should continue to perform standard Sphinx header processing and probabilistic forwarding without requiring modifications to the underlying routing protocol.

\end{description}

\color{black}{}
\section{Methodology}\label{methodology}

In this section, we first present our defense against the malicious last-mix-hop attack (Section~\ref{sec:last-hop-fix}), followed by replier-anonymous SURBs that prevent first-mix-hop header matching (Section~\ref{sec:replier-anon}). We then introduce NymHS, our hidden-service protocol for the Nym mixnet (Section~\ref{sec:hidden-service}).

\subsection{Withholding Mix-Derived Payload-Key Seeds from the Replier}
\label{sec:last-hop-fix}

Standard Nym SURBs disclose every mix-derived seed
$\sigma_i$ to the replier, allowing it to derive the per-hop
payload keys
$k_i=h_{\mathsf{k}}(\sigma_i)$.
As shown in Section~\ref{sec:first-hop-surb-attack}, these
keys enable a malicious replier that additionally controls the last mix hop of the reply path to recognize the corresponding packet through payload or shared secret matching.
However, the mix-derived seeds need not be disclosed to the replier and can instead be provided directly to the creator's gateway.
Accordingly, the replier receives only the original Sphinx SURB
format~\cite{Danezis2009Sphinx}, $\mathcal{S}^{\star}=(n_0,M_0,\tilde{k}),
$instead of Nym's extended
$(n_0,M_0,\sigma_0,\ldots,\sigma_{\nu-1},\tilde{k})$.

In addition, Nym headers support up to $r=5$ Sphinx hops, whereas ordinary
reply paths use only 4 hops, leaving approximately $60$\,B of
unused space in the final-hop routing block~$\beta$.
The creator exploits this unused capacity by embedding
\[
K_{\mathsf{emb}}
=
\sigma_0\Vert\sigma_1\Vert\sigma_2,
\qquad
|K_{\mathsf{emb}}|
=
3\kappa
=
48\,\mathrm{B},
\]
inside the final routing block and marking it with the $\textsf{EMBEDDED\_KEYS}$ flag.
The gateway derives the remaining seed
$\sigma_{\nu-1}=h_\sigma(s_{\nu-1})$
from its own shared secret.
Because $K_{\mathsf{emb}}$ resides inside the onion-encrypted
routing block, only the recipient gateway can recover it.

Importantly, under this construction, the replier no longer applies the mix-derived payload encryption.
It forms $p
=
\mathrm{SURB\text{-}ACK}
\Vert
H(\tilde{k})
\Vert
\mathrm{Enc}^{\mathsf{AES}}_{\tilde{k}}(\mathsf{FH}\Vert f)$ with the SURB-ACK and~$H(\tilde{k})$ left in cleartext outside the AES ciphertext.
Upon encountering the $\textsf{EMBEDDED\_KEYS}$ flag, the gateway recovers $K_{\mathsf{emb}}$ and uses the embedded seeds, together with its locally derived last-mix-hop seed, to reverse the mix-derived payload transformations.
It then forwards the SURB-ACK and delivers $H(\tilde{k})\Vert\mathrm{Enc}^{\mathsf{AES}}_{\tilde{k}}(\mathsf{FH}\Vert f)$ to the client, which matches the digest to the corresponding key and decrypts the AES ciphertext.
This defense leaves the Sphinx packet format and header size unchanged while eliminating both last-mix-hop payload matching and seed matching.\footnote{The hybrid games $G$--$G_3$ of Danezis and Goldberg~\cite{Danezis2009Sphinx} continue to apply.}
The complete equations are provided in Appendix~\ref{app:last-hop-keys}.\footnote{Section~\ref{sec:replier-anon} uses $\mathcal{S}^{\star}$ as the inner SURB of a rewrapped reply.}

\subsection{Extending SURBs with Replier Anonymity}
\label{sec:replier-anon}

As discussed earlier, standard SURBs do not provide replier anonymity because $M_0$ enters the mixnet directly from the replier's gateway. Consequently, a malicious SURB creator that controls the first mixnode on the reply route can recognize the corresponding header and infer the replier's long-term gateway. To prevent this attack, the replier must avoid injecting the reply directly onto the precomputed SURB route. We therefore conceal $M_0$ until a designated \emph{swap hop} by carrying it inside an independently routed outer Sphinx packet. The outer packet first traverses a separate route to the swap node, which then transfers the packet onto the original reply route. As a result, the first mixnode of the inner route observes the packet arriving from the swap node rather than from the replier's gateway, satisfying Requirements~\textbf{R1} and~\textbf{R6}.

Carrying the inner header increases the payload size from~$\ell$ to~$\ell+|M|$. We additionally replace the Sphinx payload permutation~$\pi$ with a length-preserving pseudo one-time pad (POTP), while leaving Sphinx header processing unchanged. The inner SURB is the $\mathcal{S}^{\star}$ construction introduced in Section~\ref{sec:last-hop-fix}: the replier holds $\tilde{k}$ but does not receive the inner mix-derived seeds.\footnote{Notation follows Appendix~\ref{app:sphinx-notation}; standard SURB replies are described in Section~\ref{sec:sphinx-surb}.}

\paragraph{POTP payload encryption.}
POTP replaces $\pi_k$ (Section~\ref{sec:sphinx-surb}) with XOR against a per-hop pseudorandom stream.
With $\mathrm{PRG}_{\ell}:\{0,1\}^{\kappa}\rightarrow\{0,1\}^{\ell}$,
\begin{align}
  \mathrm{Enc}^{\mathsf{POTP}}_k(x)
    &= x \oplus \mathrm{PRG}_{|x|}(k), & 
  \mathrm{Dec}^{\mathsf{POTP}}_k(c)
    &= c \oplus \mathrm{PRG}_{|c|}(k),
    \label{eq:potp}\\
  \bigl|\mathrm{Enc}^{\mathsf{POTP}}_k(x)\bigr|
    &= |x|. \notag
\end{align}
To transport the SURB header to the swap hop, the padding procedure reserves an auxiliary suffix~$m$:
\begin{equation}
\begin{aligned}
  \mathrm{Pad}_{\mathsf{POTP}}(p,m,\ell)
  &= p \Vert \texttt{0x01} \Vert 0^{\ell-|p|-|m|-1} \Vert m,\\
  \bigl|\mathrm{Pad}_{\mathsf{POTP}}(p,m,\ell)\bigr|
  &= \ell.
\end{aligned}
\label{eq:pad-potp}
\end{equation}

\paragraph{Inner reply construction.}
Let $\mathcal{S}^{\star}=(n_0,M_0,\tilde{k})$ with $M_0=(\alpha_0,\beta_0,\gamma_0)$ as in Section~\ref{sec:last-hop-fix}.
Write $m_0$ for the $|M|$-byte encoding of $M_0$.
For fragment $f_i$, following Section~\ref{sec:surb-transmission}, the replier forms
\[
p_i
=
\mathrm{SURB\text{-}ACK}_i
\Vert
H(\tilde{k})
\Vert
\mathrm{Enc}^{\mathsf{AES}}_{\tilde{k}}(\mathsf{FH}_i\Vert f_i),
\]
with the SURB-ACK and~$H(\tilde{k})$ remaining in cleartext outside the AES ciphertext.
Unlike the construction in Section~\ref{sec:surb-transmission}, the replier does \emph{not} apply the inner mix keys
$k_0^{\mathsf{POTP}},\ldots,k_{\nu-1}^{\mathsf{POTP}}$
because the corresponding seeds are embedded in $\beta_{\nu-1}$ as $K_{\mathsf{emb}}$.
Instead, it pads the reply into the extended payload field while embedding the inner header~$m_0$:
\begin{equation}
\delta_0
=
\mathrm{Pad}_{\mathsf{POTP}}(p_i,m_0,\ell+|M|),
\qquad
|\delta_0|=\ell+|M|,
\qquad
\operatorname{tail}_{|M|}(\delta_0)=m_0.
\label{eq:delta-inner-tail}
\end{equation}

\paragraph{Outer rewrap and header swap.}
The replier constructs its own fresh Sphinx header~$M_0'$ on a route $(n_0',\ldots,n_{D-1}')$ of depth~$D$ that it selects independently of the inner route, and encapsulates $\delta_0$ in the resulting outer packet $(M_0',\delta_0')$; unlike the inner header~$M_0$, which the creator precomputed and supplied in the SURB, $M_0'$ is generated by the replier. The final outer hop~$n_{D-1}'$ is marked with $\textsf{SWAP\_HEADER}$, and the inner first mix hop~$n_0$ is embedded in $\beta_{D-1}'$ (Appendix~\ref{app:sphinx-header}).
Outer seeds/keys satisfy $\sigma_j'=h_\sigma(s_j')$ and $k_j^{\mathsf{POTP}\prime}=h_{\mathsf{k}}(\sigma_j')$:
\begin{equation}
  \delta_0'
  =
  \mathrm{Enc}^{\mathsf{POTP}}_{k_0^{\mathsf{POTP}\prime}}
  \Bigl(
  \cdots
  \mathrm{Enc}^{\mathsf{POTP}}_{k_{D-1}^{\mathsf{POTP}\prime}}
  (\delta_0)
  \cdots
  \Bigr).
  \label{eq:delta-outer}
\end{equation}
Each outer hop processes the Sphinx header and removes its corresponding POTP layer as usual.
At $n_{D-1}'$, the $\textsf{SWAP\_HEADER}$ flag instructs the node to recover $n_0$ from $\beta_{D-1}'$, remove the final outer POTP layer to obtain $\delta_0$, and split
\begin{align}
m_0 &= \operatorname{tail}_{|M|}(\delta_0), \label{eq:swap-extract}\\
\delta_{\mathrm{body}}
&=
\operatorname{head}_{\ell}(\delta_0),
\end{align}
parsing $m_0$ as $M_0$.
The node then replaces the extracted header bytes with fresh randomness $R\xleftarrow{\$}\{0,1\}^{|M|}$, forming $\delta_0^\star=\delta_{\mathrm{body}}\Vert R$, and forwards $(M_0,\delta_0^\star)$ to~$n_0$.

From this point onward, the packet follows the inner path described in Section~\ref{sec:last-hop-fix}: each mixnode removes its locally derived payload layer, and the gateway recovers $K_{\mathsf{emb}}$ upon processing the $\textsf{EMBEDDED\_KEYS}$ flag. Consequently, $n_0$ receives $M_0$ from the swap node rather than directly from the replier's gateway, defeating the first-mix-hop attack described in Section~\ref{sec:first-hop-surb-attack}. At the same time, the replier never obtains the inner mix-derived seeds required to perform last-mix-hop matching. The resulting SURB representation is also $64$ bytes smaller.

\paragraph{Security of the first-mix-hop defense.}
\label{par:replier-anon-security}
POTP preserves the payload length while replacing the Sphinx payload permutation with a pseudorandom one-time pad. Under the security of the underlying PRG, the resulting ciphertext is computationally indistinguishable from a uniformly random string of the same length. Moreover, because $M_0$ remains concealed until the swap hop emits $(M_0,\delta_0^\star)$, a malicious SURB creator controlling~$n_0$ cannot associate the observed header with the replier's gateway. Rewrap-and-swap slightly relaxes the stratified routing structure of Nym~\cite{diaz2021nym}, as the swap node may forward a packet from a later layer of the outer route to the first layer of the inner route. Table~\ref{tab:surb-comparison} summarizes the resulting differences between standard and replier-anonymous SURBs.
\color{black}{
\subsection{The NymHS Hidden-Service Protocol}
\label{sec:hidden-service}

Building on the replier-anonymous SURBs developed above, we introduce \emph{NymHS}, a hidden-service protocol for the Nym mixnet that enables mutually anonymous communication between clients and services. NymHS combines authenticated SURB publication, anonymous service discovery, pseudonymous session establishment, asynchronous bidirectional communication, and anonymous SURB replenishment through a rendezvous layer of repository nodes. Specifically, repositories store precomputed SURBs but do not participate in end-to-end communication. Hidden services periodically publish authenticated SURBs to a deterministically selected set of repositories, allowing clients to anonymously retrieve fresh reply paths without revealing the service's network location. All communication among clients, services, and repositories is carried in Sphinx packets routed through the mixnet. Reply and acknowledgment packets use the replier-anonymous SURBs introduced in Section~\ref{sec:replier-anon}, thereby satisfying Requirements~\textbf{R1} and~\textbf{R6}.

At the application layer, messages carry an Ed25519 $\mathsf{senderTag}$, a message $\mathsf{kind}$, a signature~$\sigma$, and a fresh $\mathsf{nonce}$ to provide authentication, integrity, and replay protection (Table~\ref{tab:hs-message}). The prototype binaries and $\mathsf{kind}$ dispatch mechanism are summarized in Appendix~\ref{app:implementation}.
\subsection*{Roles in the Protocol}

The NymHS protocol involves the following roles:

\begin{itemize}
    \item \textbf{Client:} An entity that initiates communication with a hidden service while concealing its own network location.
    \item \textbf{Hidden Service:} A service that receives and responds to client requests while concealing its network location.
    \item \textbf{Repository Nodes:} A set of nodes responsible for storing and forwarding signed SURBs published by hidden services, as well as encrypted SURBs exchanged by clients and hidden services.
\end{itemize}
All communication among these entities is conducted exclusively through the mixnet.
\paragraph{Terminology.}
Creator and replier name roles on a single SURB and cut across the roles above.
Either party can hold either: a client creates the SURBs on which it expects replies, whereas a hidden service creates the SURBs that clients use to reach it.
Because a SURB returns to whoever created it, its destination is likewise a client in the first case and a service in the second.

\subsection*{SURB Publication and Verification}
\label{sec:hs-publication}

Before clients can initiate communication, hidden services publish fresh SURBs to the repository layer. The publication process proceeds as follows:

\begin{enumerate}
\item The service computes $h(PK_S \,\|\, \mathit{epoch})$, where $\mathit{epoch}$ denotes the current network-wide epoch, i.e., a fixed-duration interval defined by the Nym protocol during which the network topology remains stable before being refreshed~\cite{diaz2021nym}. The resulting digest is used to select the $N$ closest repository nodes for SURB storage.
\item For each selected repository node $\mathcal{R}_i$, the service sends a message through the mixnet containing:
\begin{enumerate}[label=(\roman*)]
\item the service public key $PK_S$;
        \item a large secret value $\mathcal{U}$ shared between the service and $\mathcal{R}_i$;
        \item a set of signed service SURBs $\mathcal{S}_S = \{(\mathcal{S}_j, \text{sig}_j)\}_{j=1}^n$;
        \item a set of auxiliary service SURBs $\mathcal{S}^{\mathsf{aux}}_S = \{\mathcal{S}^{\mathsf{aux}}_{S,k}\}_{k=1}^m$, used for acknowledgments and replenishment requests.
    \end{enumerate}
    \item Upon receiving the publication, $\mathcal{R}_i$ verifies the signatures. If verification succeeds, the repository stores the SURBs in a data structure indexed by the service public key $PK_S$ and returns an acknowledgment using one auxiliary service SURB from $\mathcal{S}^{\mathsf{aux}}_S$. Otherwise, the publication is discarded and an error is returned.
\end{enumerate}

Because repositories index published SURBs by $PK_S$, the service public key effectively serves as a stable, location-independent address. Any client that knows $PK_S$ can derive the same repository set, retrieve fresh reply paths, and contact the service without learning its gateway or underlying network location.

\subsection*{SURB Retrieval by the Client}
\label{sec:hs-retrieval}

Once a hidden service has published its SURBs, clients can retrieve them anonymously through the repository layer:

\begin{enumerate}
    \item The client computes $h(PK_S \,\|\, \text{epoch})$ to identify the same set of $N$ repository nodes selected by the service.
    \item The client randomly selects one repository node $\mathcal{R}_i$ and sends a retrieval request through the mixnet, including the service public key $PK_S$ and a set of auxiliary client SURBs $\mathcal{S}^{\mathsf{aux}}_C = \{\mathcal{S}^{\mathsf{aux}}_{C,j}\}_{j=1}^{m'}$ for receiving the response.
    \item If at least $k$ valid service SURBs are available, $\mathcal{R}_i$ returns them to the client using SURBs from $\mathcal{S}^{\mathsf{aux}}_C$. The client verifies the received SURBs and stores them locally.
    \item If fewer than $k$ SURBs are available:
\begin{enumerate}[label=(\roman*)]
        \item the repository node returns an error to the client using an auxiliary client SURB;
        \item if the service is already known to the repository, the repository initiates a replenishment request to the service using auxiliary service SURBs from $\mathcal{S}^{\mathsf{aux}}_S$ and the shared secret $\mathcal{U}$;
        \item the client may retry the retrieval after replenishment or contact another repository node from the derived repository set.
    \end{enumerate}
\end{enumerate}

The client stores the retrieved SURBs in a local data structure indexed by the hidden service public key.

\subsection*{Establishing Communication with the Hidden Service}
\label{sec:hs-session}

Once the client has obtained a sufficient number of valid SURBs, it can initiate anonymous communication with the hidden service. At the beginning of each session, the client generates an ephemeral key pair $(sk_C^{\mathsf{sess}}, pk_C^{\mathsf{sess}})$. The public key $pk_C^{\mathsf{sess}}$ serves as a session-scoped pseudonym in messages sent to the service. It does not reveal the client's network identity; instead, the service uses it to index session-specific state, including stored SURBs, and to verify incoming packets.

Communication proceeds as follows:

\begin{enumerate}
\item The client initiates communication by transmitting authenticated web requests using the retrieved service SURBs (Section~\ref{methodology}) and includes its session public key $pk_C^{\mathsf{sess}}$. The hidden service uses this key to authenticate requests and associate them with the corresponding pseudonymous session. The initial request carries a larger set of \emph{bootstrap SURBs} $\mathcal{S}^{C}_{\mathsf{boot}}$, where $|\mathcal{S}^{C}_{\mathsf{boot}}| \gg |\mathcal{S}^{C}_{\mathsf{cont}}|$, to establish sufficient reply capacity. Each subsequent request carries a smaller set of \emph{continuation SURBs} $\mathcal{S}^{C}_{\mathsf{cont}}$ to replenish the reply paths consumed during the session.

\item The hidden service stores the received SURBs indexed by $pk_C^{\mathsf{sess}}$, forwards authenticated requests to a local application endpoint, and returns signed responses through the stored SURBs using replier-anonymous replies (Section~\ref{sec:replier-anon}). Each response additionally carries fresh reply SURBs $\mathcal{S}_{\mathsf{server}}$, allowing the client to continue communicating without exhausting its available reply paths.

\item Both the client and the service reserve a fixed threshold of SURBs, denoted by $\tau$, exclusively for replenishment. This reserve ensures that control communication remains possible even when the general-purpose SURB pool is nearly exhausted.

\item If either party determines that transmitting a message would reduce its available SURB pool below $\tau$, the message is buffered and a replenishment protocol is initiated through a rendezvous point identified by $\mathsf{rid}$:
\begin{enumerate}[label=(\roman*)]
    \item the requester generates a Diffie--Hellman public key share $g^x$ and transmits it together with the requested number of SURBs $n_{\mathsf{req}}$, referencing the session pseudonym $pk_C^{\mathsf{sess}}$ carried in the $\mathsf{senderTag}$ field (Table~\ref{tab:hs-message});
    \item the peer generates its own Diffie--Hellman share $g^y$, derives the shared secret $K=g^{xy}$, and generates a fresh set of SURBs $\mathcal{S}_{\mathsf{new}}$ with $|\mathcal{S}_{\mathsf{new}}|=n_{\mathsf{req}}+c$, where $c$ is a protocol-defined replenishment reserve. The peer encrypts the SURB set under $K$ and forwards the encrypted bundle to a selected repository node together with a rendezvous identifier $\mathsf{rid}$;
    \item after the repository node confirms storage, the peer informs the requester of $\mathsf{rid}$, the selected repository node, and the number of SURBs $n_{\mathsf{ret}}$ required to retrieve the stored bundle;
    \item The requester generates \(n_{\mathsf{ret}}+c\) SURBs, where \(c\) is a protocol constant accounting for auxiliary communication, and anonymously retrieves the encrypted SURB bundle by presenting \(\mathsf{rid}\) together with the corresponding SURBs. The repository node returns the encrypted bundle through the supplied SURBs, after which the requester decrypts \(\mathcal{S}_{\mathsf{new}}\) using \(K\) and resumes normal communication.
\end{enumerate}
\end{enumerate}

\paragraph{Design requirements.}
The protocol satisfies the design requirements defined in Section~\ref{sec:requirements} as follows. Replier-anonymous SURBs conceal the replier's gateway while preserving compatibility with the Sphinx packet format and routing semantics, thereby satisfying Requirements~\textbf{R1} and~\textbf{R6}. SURB publication and anonymous retrieval through the repository layer enable clients to discover and contact a hidden service using only its NymHS pseudonym, satisfying Requirement~\textbf{R2}. Session establishment through ephemeral pseudonymous identifiers, combined with SURB-based communication, ensures that neither endpoint reveals its network identifier or entry gateway to the other party, satisfying Requirement~\textbf{R3}. The authenticated message format together with Ed25519 signatures provides authenticated bidirectional communication, satisfying Requirement~\textbf{R4}. Finally, continuous SURB replenishment supports asynchronous communication over extended periods without requiring persistent circuits or fixed return paths, satisfying Requirement~\textbf{R5}.

\subsection{Prototype Implementation}
\label{sec:implementation}

We implemented NymHS as an extension of the Nym SDK, adding the replier-anonymous SURBs of Section~\ref{sec:replier-anon} and the hidden-service protocol of Section~\ref{sec:hidden-service} across a client proxy, a hidden service, and a repository node. Mixnodes require only the POTP payload format and the additional routing flags of Section~\ref{sec:replier-anon}; entry gateways rewrap the acknowledgments they inject, and Loopix scheduling is unchanged from Section~\ref{sec:loopix}. Appendix~\ref{app:implementation} gives the binaries, the Sphinx-layer changes, and the build parameters.%\footnote{Companion repository and README: \url{https://github.com/nionis/nym}.}
}

\color{black}{

\begin{comment}
\section{Evaluation}
\label{sec:evaluation}

Nym currently uses a $2\,\mathrm{KiB}$ Sphinx payload together with a $20\,\mathrm{ms}$ Poisson transmission interval, bounding throughput at approximately $100\,\mathrm{KiB/s}$ before protocol metadata.
At this size, roughly one fifth of each reply fragment is fixed metadata rather than application data.
Reported broadband speeds are on the order of several megabytes per second~\cite{shirvani2024past}, so the mixnet's real traffic stream---not the access link---is the throughput bottleneck.
We therefore evaluate larger payload sizes to measure how amortizing that overhead reduces fragmentation, and we vary rewrap depth to quantify the latency cost of replier anonymity.
We benchmark a single NymHS hidden service on one host using 118 static websites under a $9\times3$ design with $\ell\in\{2,3,\ldots,10\}\,\mathrm{KiB}$ and $D\in\{1,2,3\}$; each of the 27 configurations is measured three times over all 118 sites ($9{,}558$ page loads).
\end{comment}

\section{Evaluation}
\label{sec:evaluation}

In this section, we empirically evaluate NymHS. Before presenting the results, we first describe the evaluation setting and the performance characteristics that motivate our experiments.

Nym currently uses a $2\,\mathrm{KiB}$ Sphinx payload size, with client data transmitted into the mixnet according to a Poisson process with a mean inter-departure time of $20\,\mathrm{ms}$. This limits the resulting application-data throughput to approximately $100\,\mathrm{KiB/s}$. At this packet size, roughly one fifth of each reply fragment is consumed by fixed metadata and routing information, further reducing the effective payload capacity.

Moreover, reported broadband speeds are on the order of several megabytes per second~\cite{shirvani2024past}, making the mixnet's real traffic stream---rather than the client's access link---the primary throughput bottleneck. We therefore evaluate larger payload sizes to determine how amortizing this fixed overhead reduces fragmentation and improves performance. We additionally vary the rewrap depth to quantify the latency cost introduced by replier anonymity.

Under this setting, we benchmark a single NymHS hidden service on one host using 118 static websites under a $9\times3$ experimental design with $\ell\in\{2,3,\ldots,10\}\,\mathrm{KiB}$ and $D\in\{1,2,3\}$. Each of the 27 configurations is evaluated three times across all 118 websites, resulting in $9{,}558$ page loads.

\subsection{Objectives}

\begin{description}
    \item[RQ1.] How do payload size~$\ell$ and rewrap depth~$D$ affect end-to-end page-load latency?

    \item[RQ2.] What fragmentation overhead does NymHS incur per reply fragment, and to what extent does increasing payload size reduce that overhead and improve throughput?
\end{description}
\subsection{Experimental Testbed}

All experiments ran on a single virtual private server hosting the Nym mixnet and the NymHS binaries described in Section~\ref{sec:hidden-service}.
Figure~\ref{fig:eval-workflow} depicts the measurement workflow for one page-load observation.

\begin{figure}[t]
  \centering
  \input{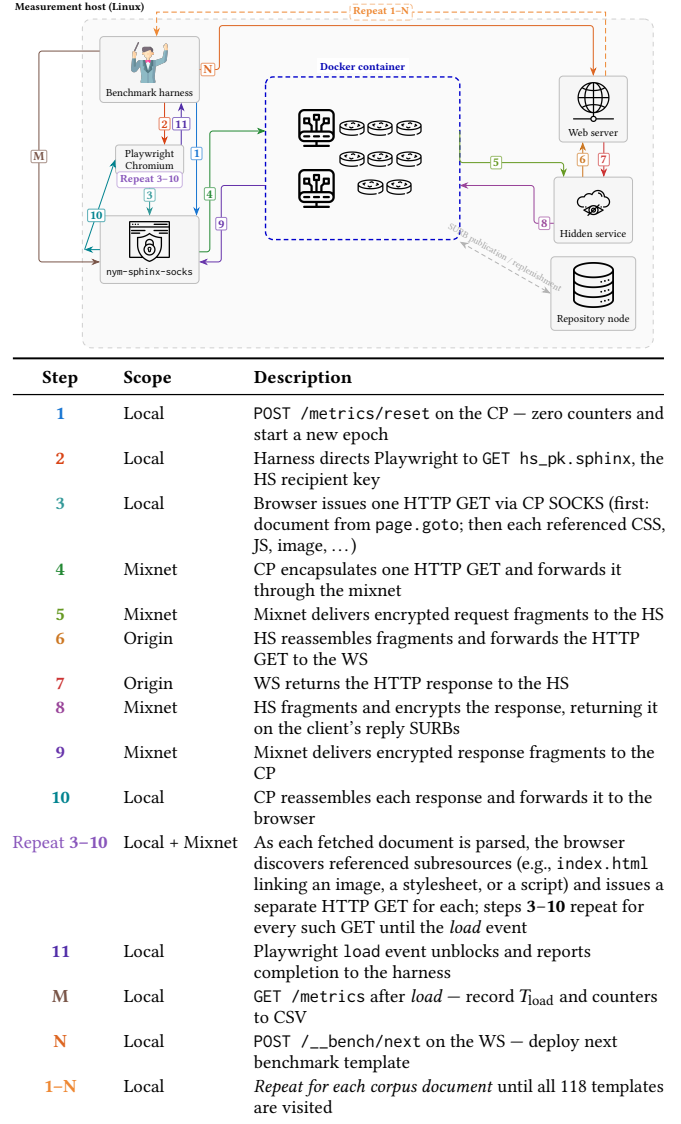}
  \caption{Evaluation workflow for one page-load observation on a single Linux server (mixnet in Docker; other components native).
  Steps \textbf{1}--\textbf{2} run once per page; \textbf{3}--\textbf{10} repeat per HTTP GET; \textbf{11} is the browser \emph{load} event; \textbf{M}/\textbf{N} record metrics and rotate the template.
  Steps \textbf{1}--\textbf{N} repeat over the $118$ corpus documents (table below). Diagram drawn with AI assistance to the authors' specification.}
  \label{fig:eval-workflow}
\end{figure}

\paragraph{Mixnet topology.}
The mixnet comprises two entry gateways, eight mixnodes in three stratified layers and a Network Requester.

\paragraph{System components.}
We deployed four NymHS processes:
\begin{enumerate}
    \item a \textbf{Repository Node}~(RN) that stores authenticated SURBs published by hidden services and coordinates SURB replenishment;
    \item a \textbf{Web Server}~(WS) hosting the benchmark site at \texttt{/}. The harness rotates templates through local \texttt{POST}~\nolinkurl{/\_\_bench/next} and \texttt{GET}~\nolinkurl{/\_\_bench/status} endpoints;
    \item a \textbf{Hidden Service}~(HS) that terminates mixnet requests, fetches content from the WS, fragments responses into Sphinx packets, and returns them via client-supplied reply SURBs;
    \item a \textbf{Client Proxy}~(CP) exposing a SOCKS interface and translating HTTP into hidden-service messages. For benchmarking, the CP exposes \texttt{POST}~\nolinkurl{/metrics/reset} and \texttt{GET}~\nolinkurl{/metrics}; counters are reset before each page load and read after the browser \emph{load} event.
\end{enumerate}

\paragraph{Benchmark harness.}
A separate harness orchestrates experiments without joining the mixnet.
For each page it resets CP counters (\texttt{POST /metrics/reset}), drives headless Chromium through Playwright~\cite{pathak2024web} with the CP SOCKS proxy, loads the HS \texttt{.sphinx} URL, waits for the browser \texttt{load} event, records metrics (\texttt{GET /metrics}), and advances the WS template (\texttt{POST /\_\_bench/next}).
Each reported epoch covers one end-to-end page retrieval.\paragraph{Rationale of the experimental setup.}
All components ran on a single host to isolate the effects of~$\ell$ and~$D$ under fixed hardware and software.
The evaluation could not use the public Nym network because NymHS changes Sphinx payload encryption (POTP) and sphinx routing (\textsf{SWAP\_HEADER,EMBEDDED\_KEYS}), which we do not control on public mix nodes or gateways; wide-area variability would also confound latency measurements.
We evaluate larger~$\ell$ rather than a shorter Poisson interval because a higher transmission rate alone would increase fragmentation, and each additional fragment would carry the same fixed SURB metadata overhead.

\subsection{Experimental Methodology}
To assess the impact of $\ell$ and $D$ on NymHS, we vary both parameters in a full-factorial design, rebuilding and restarting the stack for each of the twenty-seven configurations while holding the workload and host environment constant. Our primary outcome is the page-load latency $T_{\mathrm{load}}$.
%We varied $\ell$ and $D$ in a full factorial design, rebuilding and restarting the stack for each of the twenty-seven configurations while holding the workload and host constant.
%Primary outcome: page-load latency $T_{\mathrm{load}}$.
\subsubsection{Workload}

The workload comprises 118 static website templates (HTML with associated CSS, JavaScript, and images) obtained from a publicly available template repository.\footnote{\url{https://github.com/learning-zone/website-templates.git}}

Before benchmarking, all templates are preprocessed offline so that every resource is served locally by the WS. External assets (stylesheets, scripts, fonts, and images) are downloaded and their references rewritten to local copies. Unresolved external resources and references to missing local files are removed to prevent spurious HTTP~404 requests during evaluation. Consequently, every browser request traverses the mixnet. We exclude directories that ship only screenshots of templates and no
\texttt{index.html}, since there is no page in them to load.

Each template is loaded once per configuration through the hidden service using Playwright and the CP with browser caching disabled, and the full design is repeated three times. The host is a virtual private server whose resources are shared with other tenants, so repetition limits the influence of contention on any one run. A measurement completes when the browser reaches the \texttt{load} event or after a timeout of 600\,s.
\subsubsection{Experimental Variables}

We swept two parameters:
\begin{itemize}
    \item \textbf{Payload size ($\ell$):} $\{2,3,\ldots,10\}$~KiB.
    Replier-anonymous replies use payload field size $\ell+|M|$ with $|M|=348$\,B (Section~\ref{sec:replier-anon}).
    \item \textbf{Rewrap depth ($D$):} $\{1,2,3\}$ mixnodes on the rewrap route before header swap.
\end{itemize}
This yields a $9\times3$ design: twenty-seven configurations, each evaluated on all 118 templates and repeated three times ($9{,}558$ page-load measurements).

\subsection{Latency Evaluation}
\label{sec:eval-latency}

RQ1 asks how payload size~$\ell$ and rewrap depth~$D$ affect browsing latency.
Each reply fragment is a fixed-size packet that leaves the replier's queue under Loopix's Poisson schedule (Section~\ref{sec:loopix}), so smaller~$\ell$ forces more fragments per page.
Rewrap depth~$D$ sends every fragment through $D$ outer mix hops (Section~\ref{sec:replier-anon}), each peeling a Sphinx layer, removing a POTP layer and drawing a mix delay, but fragment count depends only on~$\ell$.

\subsubsection{Results}

At $D=1$, raising $\ell$ from $2$\,KiB to $5$\,KiB cuts mean latency from $36.06$\,s to $12.21$\,s, approximately a $3.0\times$ reduction; raising it further to $10$\,KiB reaches $6.92$\,s, $5.2\times$ faster than at $2$\,KiB.
Latency falls with every increase in~$\ell$ and rises with every increase in~$D$, with no exception anywhere in the grid~(Figure~\ref{fig:latency-vs-payload}).
Increasing $D$ from $1$ to $3$ adds between $0.36$\,s and $0.60$\,s.

\begin{table*}[t]
  \centering
  \footnotesize
  \setlength{\tabcolsep}{4pt}
  \caption{Per-fragment application budget and page-load latency $T_{\mathrm{load}}$ over 118 websites.
  Avail.\ is application bytes per fragment after fixed metadata, OH is $100\times445/\ell$, and App.\ KiB/s
  is sustained application throughput at 50 packets/s. Latencies are mean\,$\pm$\,standard deviation over
  $354$ loads per configuration ($118$ templates $\times$ $3$ repetitions). The spread is across page loads and
  reflects the size distribution of the corpus rather than measurement error
  (Appendix~\ref{app:fragments}).}
  \label{tab:latency-grid}
  \begin{tabular}{@{}lrrrccc@{}}
    \toprule
    & & & & \multicolumn{3}{c}{$T_{\mathrm{load}}$ (s), mean\,$\pm$\,sd} \\
    \cmidrule(lr){5-7}
    $\ell$ & Avail.\ (B) & OH (\%) & App.\ KiB/s & $D=1$ & $D=2$ & $D=3$ \\
    \midrule
$2$\,KiB & 1{,}603 & 21.7 & 78.3 & $36.06\pm21.70$ & $36.19\pm21.80$ & $36.66\pm21.94$ \\
$3$\,KiB & 2{,}627 & 14.5 & 128.3 & $21.03\pm12.24$ & $21.30\pm12.34$ & $21.53\pm12.39$ \\
$4$\,KiB & 3{,}651 & 10.9 & 178.3 & $15.31\pm8.61$ & $15.54\pm8.64$ & $15.77\pm8.70$ \\
$5$\,KiB & 4{,}675 & 8.7 & 228.3 & $12.21\pm6.68$ & $12.53\pm6.72$ & $12.80\pm6.81$ \\
$6$\,KiB & 5{,}699 & 7.2 & 278.3 & $10.40\pm5.49$ & $10.77\pm5.51$ & $10.94\pm5.60$ \\
$7$\,KiB & 6{,}723 & 6.2 & 328.3 & $9.10\pm4.71$ & $9.35\pm4.70$ & $9.45\pm4.77$ \\
$8$\,KiB & 7{,}747 & 5.4 & 378.3 & $8.18\pm4.10$ & $8.42\pm4.16$ & $8.64\pm4.26$ \\
$9$\,KiB & 8{,}771 & 4.8 & 428.3 & $7.40\pm3.72$ & $7.56\pm3.71$ & $7.81\pm3.89$ \\
$10$\,KiB & 9{,}795 & 4.3 & 478.3 & $6.92\pm3.35$ & $7.05\pm3.45$ & $7.37\pm3.48$ \\
    \bottomrule
  \end{tabular}
\end{table*}

\begin{figure}[t]
  \centering
  \includegraphics[width=\columnwidth]{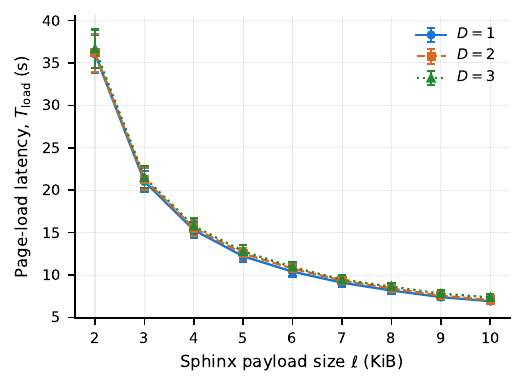}
  \caption{Mean page-load latency $T_{\mathrm{load}}$ versus payload size~$\ell$ for rewrap depths $D\in\{1,2,3\}$.}
  \label{fig:latency-vs-payload}
\end{figure}

%\subsubsection{Discussion}

\paragraph{Throughput bound.}
The Poisson schedule of Section~\ref{sec:loopix} has a $20\,\mathrm{ms}$ mean interval, so a client sends approximately $50$ packets per second whatever the payload size.
Each packet carries $\ell-445$\,B of application data, where the $445$\,B of fixed metadata is itemised in Section~\ref{sec:eval-overhead}, bounding sustained throughput at $50(\ell-445)$\,B/s: the \emph{App.\ KiB/s} column of Table~\ref{tab:latency-grid}, from $78.3\,\mathrm{KiB/s}$ at $\ell=2\,\mathrm{KiB}$ to $478.3\,\mathrm{KiB/s}$ at $\ell=10\,\mathrm{KiB}$.
Since the packet rate is fixed, the only way to send more application data is to put more of it in each packet, which is why the $445$\,B of fixed metadata costs $21.7\%$ of a $2$\,KiB packet but only $4.3\%$ of a $10$\,KiB one.
\paragraph{Why load times vary so widely.}
A page's latency is set by how many fragments it needs.
At $\ell=2$\,KiB each fragment carries $1{,}603$\,B of application data. Each response is fragmented as its own set (Section~\ref{sec:surb-transmission}). A page of $B$ bytes delivered in a single response therefore needs $\lceil B/1{,}603\rceil$ fragments, and a multi-response page somewhat more.
The smallest page is a single $2{,}416$\,B response needing two fragments, and it loads in $2.3$\,s; the largest needs at least $2{,}953$ and takes $98$\,s. Raising $\ell$ to $10$\,KiB cuts that same page to at least $484$ fragments and $16$\,s.
The standard deviations reported in Table~\ref{tab:latency-grid} are therefore closely related to the number of fragments required to load each page: larger pages require substantially more fragments and consequently exhibit higher mean load times. Appendix~\ref{app:fragments} shows that page-load latency increases approximately linearly with fragment count, to the extent that page size alone provides a strong predictor of load time.
This relationship is driven by two main components. First, each page load incurs a per-fragment transmission cost, whose total contribution scales with the number of fragments required. Second, it incurs per-page overhead that is largely independent of the amount of application data transferred, most notably the replenishment protocol, during which reply SURBs must first be deposited at a repository node and subsequently retrieved.
Raising $\ell$ from $2$ to $10$\,KiB multiplies the application data per fragment by $6.1$ and scales the transmission term with it, but because of other factors like the replenishment protocol, latency falls only $5.2\times$.
For deployment, $\ell=10$\,KiB is the fastest setting, and since $D$ costs well under a second at any~$\ell$ it can be set by the anonymity wanted before the header swap rather than by speed.

\subsection{Communication Overhead}
\label{sec:eval-overhead}

RQ2 quantifies the fragmentation and protocol overhead of NymHS reply packets and shows how increasing payload size amortizes that overhead.

\paragraph{Fragmentation overhead and available application data.}
Replier-anonymous replies (Section~\ref{sec:replier-anon}) use a payload field of size $\ell+|M|$, where $|M|=348$\,B is the withheld Sphinx header. Of this capacity, every fragment reserves fixed space for protocol metadata (Section~\ref{sec:surb-transmission}): a $\mathrm{SURB\text{-}ACK}$ ($405$\,B), $H(\tilde{k})$ ($32$\,B), the fragment header $\mathsf{FH}_i$ ($7$\,B), the one-byte $\mathrm{Pad}_{POTP}$ separator, and the embedded header $M$. The remaining application payload per fragment is therefore
\[
(\ell+|M|)-|\mathrm{SURB\text{-}ACK}|-|H(\tilde{k})|-|\mathsf{FH}_i|-|M|-1
= \ell-445~\text{B}.
\]
The relative protocol overhead is therefore $445/\ell$; the OH (\%) column in Table~\ref{tab:latency-grid} reports $100\times445/\ell$, and App.\ KiB/s is the corresponding sustained application throughput at 50 packets/s.
As payload size increases, the fixed overhead is amortized across larger fragments, reducing the relative overhead from $21.7\%$ to $4.3\%$ and raising application throughput from $78.3$ to $478.3$\,KiB/s.

\paragraph{Limitations.}
The corpus is drawn from a single public template repository and consists of static pages, so it does not exercise dynamic content. Our evaluation further assumes that packets are delivered without loss. The measured benefits of larger payloads therefore reflect reduced fragmentation under reliable delivery; packet loss could offset these gains by increasing retransmissions and recovery delay. In addition, all experiments were conducted in a controlled single-host environment, so the results do not capture wide-area effects such as inter-country routing, heterogeneous network paths, or Internet congestion. Evaluating NymHS over geographically distributed deployments is therefore necessary to quantify its end-to-end latency under realistic network conditions.}

\section{Related Work}
\label{relatedwork}

\color{black}{

This section situates NymHS among prior anonymous communication systems.
Section~\ref{sec:rw-routing} compares onion-routing and mixnet architectures that offer or omit hidden services; Section~\ref{sec:rw-messaging} discusses metadata-private messaging systems that address a different communication model; and Section~\ref{sec:rw-nym-apps} examines applications built on Nym, including Pudding~\cite{Kocaogullar2024Pudding}, and how the malicious first-last-mix-hop SURB attacks affects their security claims.

\subsection{Anonymous Routing Systems}
\label{sec:rw-routing}

\paragraph{Sphinx and nymservers.}
Sphinx~\cite{Danezis2009Sphinx} introduced compact layered routing with constant-size packets, per-hop unlinkability, and SURBs for anonymous replies.
It also proposes depositing SURBs at a \emph{nymserver}~\cite{Mazieres1998Nymserver} that holds a public pseudonym and injects inbound messages via stored SURBs without revealing which network identity owns the pseudonym~\cite{Mazieres1998Nymserver,Danezis2009Sphinx}.
Nymservers were conceived primarily to send and receive electronic mail under a pseudonym~\cite{danezis2003mixminion}.
Sphinx remains the packet format used by Loopix and Nym, but nymservers alone provide neither mutually anonymous client--service rendezvous nor discoverable hidden services.
\paragraph{Onion routing and Tor.}
Tor forwards traffic over persistent \emph{circuits}: fixed relay sequences with FIFO forwarding and layered onion encryption, so each relay learns only its predecessor and successor \cite{syverson2004tor}.
A Tor hidden service keeps long-lived circuits, typically refreshed every 18--24 hours, to a small set of relays called \emph{introduction points}, and publishes those points in a descriptor signed under the service's long-term public key \cite{platzer2020critical,constantinides2026ve}.
A client fetches the descriptor from Hidden Service Directories (HSDirs), builds a circuit to a chosen \emph{rendezvous point}, and uses an introduction point to ask the service to join that rendezvous over a separate circuit, so neither party reveals its network location.
Tor is vulnerable to end-to-end correlation attacks that can deanonymize a connection \cite{johnson2013users,diaz2021nym}. Tor provides no built-in Sybil-resistance mechanism, allowing any party to operate relays. This has made adversarial relay operation a realistic threat, with reports of intelligence agencies operating Tor relays for traffic collection and deanonymization attempts~\cite{ccalicskan2015technical}.

\paragraph{I2P.}
The Invisible Internet Project (I2P) is a peer-to-peer anonymous network in which every participant is both client and relay~\cite{hoang2018empirical}.
Unlike Tor's bidirectional circuits, I2P uses short-lived unidirectional tunnels and \emph{garlic routing}~\cite{diaz2021nym}, and discovers peers through a distributed hash table (DHT) rather than directory authorities~\cite{troncoso2017systematizing}.
It supports hidden services (\emph{eepsites}) for intra-network communication~\cite{hoang2018empirical} but remains vulnerable to censorship, eclipse, and deanonymization attacks~\cite{egger2013practical,herrmann2011privacy,hoang2018empirical,jeong2016longitudinal}, and---like Tor---targets a local rather than global adversary~\cite{diaz2021nym}.

\paragraph{Comparison with Nym}
Although Tor and I2P provide hidden services, they target a local adversary and weaken when both communication ends (or a large network fraction) are observed.
Nym's Loopix design instead uses independently routed Sphinx packets, probabilistic delays, and continuous cover traffic against a passive global adversary, with stake-based Sybil resistance~\cite{diaz2021nym,diaz2022reward}.
Prior to this work Nym lacked a hidden-service protocol; NymHS fills that gap while preventing the malicious-mix-hop SURB attacks.
Table~\ref{tab:tor-vs-nym} summarizes these differences.
The final row reflects Nym prior to this work: NymHS adds hidden-service support while retaining the mixnet properties above.

\begin{table*}[t]
\small
\setlength{\tabcolsep}{4pt}
\renewcommand{\arraystretch}{0.92}
\centering
\caption{Architectural comparison between Tor, I2P, and the Nym mixnet.}
\label{tab:tor-vs-nym}
\begin{tabular}{|p{0.28\textwidth}|c|c|c|}
\hline
\textbf{Property} & \textbf{Tor} & \textbf{I2P} & \textbf{Nym} \\
\hline
Architecture type & Onion routing & Peer-to-peer garlic routing & Mixnet (Loopix) \\
\hline
Communication model & Persistent circuits & Unidirectional tunnels & Independent routing \\
\hline
Packet ordering preserved & \cmark & \cmark & \xmark \\
\hline
Probabilistic forwarding delays & \xmark & \xmark & \cmark \\
\hline
Continuous loop cover traffic & \xmark & \xmark & \cmark \\
\hline
End-to-end-correlation resistance & \xmark & \xmark & Stronger \\
\hline
Website-fingerprinting resistance & \xmark & \xmark & Stronger \\
\hline
Passive global adversary model & \xmark & \xmark & \cmark \\
\hline
Sybil-resistance incentives & \xmark & \xmark & \cmark \\
\hline
Unrestricted relay participation & \cmark & \cmark & \xmark \\
\hline
Hidden-service support & \cmark & \cmark & \xmark \\
\hline
\end{tabular}
\end{table*}

\subsection{Anonymous Messaging Systems}
\label{sec:rw-messaging}
\paragraph{Kerblam.}
Kerblam~\cite{jia2025kerblam} is a two-server anonymous messaging system that provides \emph{end-to-end unlinkability}, ensuring that neither the sender nor the receiver learns the other's identity. It achieves this through oblivious message retrieval, oblivious shuffling, and a specialized ORAM$^{-}$ construction for efficient asynchronous retrieval of large messages. In contrast, NymHS builds on a decentralized Loopix/Nym mixnet rather than two non-colluding servers. While Kerblam focuses on mutually anonymous messaging, NymHS additionally supports hidden services, enabling anonymous client--server communication for web applications in which clients can issue requests and receive responses without revealing either endpoint's network location.

\paragraph{Stadium.}
Stadium~\cite{tyagi2017stadium} is a metadata-private messaging system that protects both message contents and communication metadata against a global adversary using differential privacy, cover traffic, and a distributed verifiable mixnet. It assumes that communicating parties already know one another's identifiers and focuses on concealing the communication relationship between them rather than enabling anonymous service discovery or hidden services. In contrast, NymHS provides mutually anonymous client--server communication through hidden services, allowing clients to discover and communicate with services using only public pseudonyms while hiding the network locations of both endpoints.

\subsection{Applications Built on the Nym Mixnet}
\label{sec:rw-nym-apps}
\paragraph{Clearnet browsing via exit gateways.}
To reach the public Internet a Nym client does not dial the destination itself: its entry gateway injects Sphinx packets toward an \emph{exit gateway}, which terminates the anonymous path, opens the TCP/TLS session, and answers through client-supplied SURBs.
The exit necessarily learns the destination; what the mixnet must keep from it is which entry gateway originated the request, and so the client's long-term attachment point.
Answering on SURBs, however, makes the exit a replier, so both attacks of Section~\ref{sec:first-hop-surb-attack} are available to it and both disclose exactly that gateway, reducing deanonymization to compromising or flooding one gateway rather than breaking the full mix path.

\paragraph{Nym service providers.}
Nym applications communicate through stable \emph{Nym addresses}. A provider registers identity and encryption keys with an entry gateway, yielding $\mathit{identity}.\mathit{encryption}@\mathit{gateway}$. A client sends its messages, together with the SURBs it expects answers on, over a route whose last hop is that gateway; the gateway resolves the identity field to the attached party and delivers over a persistent connection.
A malicious service provider can leverage the attacks of Section~\ref{sec:first-hop-surb-attack}: a client that withholds its own address and supplies only SURBs is the creator and the provider the replier, so the attacks expose the client's gateway, the one attachment point still concealed.

Both deployments are repaired without redesign; Figures~\ref{fig:clearnet-usecase} and~\ref{fig:nymaddr-usecase} in Appendix~\ref{app:usecases} show the two configurations. The client supplies $\mathcal{S}^{\star}$ in place of a seed-carrying SURB (Section~\ref{sec:last-hop-fix}) and its gateway rewraps the acknowledgment it injects (Section~\ref{sec:replier-anon}), so neither the last mix hop of the reply nor the first mix hop of the acknowledgment reveals where the client attaches.

\paragraph{Pudding.}
Pudding~\cite{Kocaogullar2024Pudding} provides private user discovery on Loopix/Nym: a client contacts a peer knowing only a short username such as an email address.
Users register with discovery servers, proving ownership of that identifier and storing their Nym contact information under it; a querier looks the username up at several servers, which return a matching \emph{deterministically generated SURB} (Appendix~\ref{app:pudding-surb}), and sends a first contact message through it via a \emph{reflector}.
Two of its goals concern us: \textbf{G1} (\emph{unlinkability}), that lookups of the same target by different honest users be indistinguishable, and \textbf{G4} (\emph{membership unobservability}), that outsiders cannot tell whether a username is registered; \textbf{G2} and \textbf{G3} cover impersonation resistance and binding to a real-world identifier, which our attack does not touch.
Unregistered usernames still yield a SURB, addressed to a sink, so the reply format alone does not reveal membership.
These goals cover discovery rather than long-lived mutually anonymous sessions, and a discovery server holds the reachability information~$\Delta$ registered under each username, and with it that user's entry gateway.

\paragraph{Security implications for Pudding on Nym.}
Pudding is aware that a discovery-issued SURB header is recognizable, and routes the first contact packet through a reflector so that the known header enters the mixnet away from the querier. It does not address the $\mathrm{SURB\text{-}ACK}$ variant of the attack (Section~\ref{sec:first-hop-surb-attack}).
A discovery server placed first on a $\mathrm{SURB\text{-}ACK}$ path can therefore distinguish two users who look up the same target but attach through different entry gateways, breaking \textbf{G1}, and deanonymizes one outright if that gateway is also compromised.
\textbf{G4} assumes mix, provider and discovery nodes are honest, and holds under that model. Their prototype, however, derives the fake contact information from a real mixnet client and reuses that single address for every unregistered lookup, so it names a real entry gateway rather than the nonexistent sink the paper describes.
Once the adversary is allowed a mix node, one that recognizes the deterministically generated header sees that same gateway recur across every non-member query while registered lookups disperse across the gateways their owners use, which separates registered from unregistered identifiers more readily than the published design would allow.

%\subsection{Research Gap}

%Sphinx provides anonymous packet forwarding and reply blocks but not mutually anonymous hidden services. Tor provides mature hidden-service support, but relies on persistent circuits and targets a different adversary model. Nym provides a production Loopix-based mixnet and application SDK but inherits the malicious first-hop SURB attack of Section~\ref{sec:first-hop-surb-attack} and therefore does not securely support hidden services. Pudding demonstrates private service discovery on Nym but assumes protocol conditions that exclude this attack and does not provide long-lived mutually anonymous hidden-service communication.

%These limitations leave an open problem: enabling authenticated, long-lived hidden-service communication on Loopix/Nym while preserving replier anonymity against malicious first-hop SURB attacks. NymHS addresses this problem by strengthening SURBs and building authenticated service publication, anonymous SURB retrieval, and replenishment on top of them.

}

\color{black}{}

\section{Conclusion}

Although classical mixnet protocols provide strong sender anonymity, they generally lack mechanisms for simultaneously protecting both sender and receiver privacy. Addressing this gap, we presented the first hidden-service protocol for the Nym mixnet. We first showed that naively using standard Sphinx SURBs to provide receiver anonymity introduces two practical attacks in which either a malicious SURB creator or a malicious replier, while additionally controlling only a single mixnode, can expose the other party's long-term gateway. We then developed defenses against both the first- and last-mix-hop attacks.

Building on these defenses, we introduced \emph{NymHS}, which provides simultaneous sender and receiver anonymity through authenticated SURB publication, anonymous service discovery, pseudonymous sessions, asynchronous bidirectional communication, and SURB replenishment. NymHS requires only limited changes to the underlying mixnet, namely support for the POTP payload format and three additional routing flags.
We implemented NymHS on the Nym codebase and evaluated it across $118$ websites and $9{,}558$ page loads. Our results show that reply fragmentation is the dominant factor governing page-load latency: reducing the payload size from $10$ to $2$\,KiB increases mean page-load latency by $29.1$\,s. These results demonstrate that, with appropriately sized payloads, mutually anonymous hidden services can be implemented practically over mixnets, paving the way for deployable receiver-private services in Nym and related mixnet architectures.

\begin{ethics}
All experiments were carried out on a private mixnet running in Docker on a
single host under our control, using 118 static website templates from a public
repository served locally. No human subjects were involved, no personal data was
collected, and no traffic was ever sent to, observed on, or injected into the
live Nym network. We did not use the attacks to deanonymize the gateway of any
real user. We notified the Nym developers, and the defenses of
Sections~\ref{sec:last-hop-fix} and~\ref{sec:replier-anon} are published
alongside the attacks so that the mitigations are available as the
vulnerabilities become public.
\end{ethics}

\begin{openscience}
    We release an artifact containing both the prototype and the measurements behind
  Section~\ref{sec:evaluation}. The implementation comprises the hidden service, the
  repository node and the \texttt{.sphinx} SOCKS proxy, the Sphinx-layer changes of 
  Sections~\ref{sec:last-hop-fix} and~\ref{sec:replier-anon}, and the containerised test
  network on which the measurements were taken. The data comprises the $9{,}558$ page
  loads of the $9\times3$ design, one CSV per configuration and repetition, together 
  with three analysis scripts: \texttt{verify.py} recomputes each quantitative claim and
  prints the value stated in the paper beside the value computed from the data; 
  \texttt{make\_table.py} regenerates the body of Table~\ref{tab:latency-grid}; and
  \texttt{make\_figures.py} regenerates Figures~\ref{fig:latency-vs-payload},
  \ref{fig:latency-vs-fragments} and~\ref{fig:latency-vs-fragments-avg}.
  
  Reproducing the reported numbers requires only Python and the included CSVs: no
  mixnet, no container runtime and no compilation. Reproducing the \emph{measurements}
  requires Rust and a host able to run the test network, on which one sweep of the $27$
  configurations takes on the order of a day. The artifact is
  available at \url{https://github.com/nionis/nym}. A short video showing the
hidden service being reached through a browser is at
\url{https://www.youtube.com/watch?v=EcyajkqzK5k}.

\end{openscience}

\begin{ai}
AI-based tools were used in the preparation of this work, as described below.

\emph{Writing.} The authors used generative AI-based tools (ChatGPT, Claude, and
Cursor) to revise the text, improve flow, and correct typos, grammatical errors,
and awkward phrasing.
\emph{Prototype implementation.} The prototypes were initially written for a
master's thesis. No AI-based tool generated code at that stage; ChatGPT's search
was used only to locate library documentation and to confirm Rust syntax. Cursor
and Claude were then used to rewrite the implementation in full, from the
beginning, to make it cleaner and better structured. The authors debugged the
resulting implementation manually and verified its correctness, checking every
part of the prototype by hand.
\emph{Data analysis.} Cursor and Claude were used to assist with the analysis of
the experimental data, and Cursor was used to assist in building the analysis
pipeline.
\emph{Figures.} Figures~\ref{fig:sphinx-filler} and~\ref{fig:sphinx-header} were
drawn with the assistance of Cursor. Figures~\ref{fig:sphinx-header-embedded},
\ref{fig:clearnet-usecase}, \ref{fig:nymaddr-usecase},
and~\ref{fig:eval-workflow} were drawn with the assistance of Cursor and Claude.
For each figure the authors specified the content, structure, and every element
to be depicted, and verified the rendered result; the tools were used to realise
those specifications.

We have manually verified and are responsible for the accuracy, originality, and
integrity of the output of all AI-based tools.
\end{ai}

% The next two lines define the bibliography style to be used (do not change),
% and the bibliography file (use as many or few as you wish).
\bibliographystyle{ACM-Reference-Format}
\bibliography{sample-base}

@inproceedings{Danezis2009Sphinx,
author    = {George Danezis and Ian Goldberg},
title     = {Sphinx: A Compact and Provably Secure Mix Format},
booktitle = {Proceedings of the 30th IEEE Symposium on Security and Privacy (SP 2009)},
year      = {2009},
pages     = {269--282},
publisher = {IEEE},
doi       = {10.1109/SP.2009.15}
}

@misc{Diaz2021Nym,
author       = {Claudia Diaz and Harry Halpin and Aggelos Kiayias},
title        = {The Nym Network: The Next Generation of Privacy Infrastructure},
year         = {2021},
howpublished = {White Paper},
institution  = {Nym Technologies SA},
note         = {Version 1.0, February 26, 2021},
url          = {https://nym.com/nym-whitepaper.pdf}
}

@inproceedings{Kocaogullar2024Pudding,
author    = {Ceren Kocao\u{g}ullar and Daniel Hugenroth and Martin Kleppmann and Alastair R. Beresford},
title     = {Pudding: Private User Discovery in Anonymity Networks},
booktitle = {Proceedings of the IEEE Symposium on Security and Privacy (SP)},
year      = {2024},
pages      = {3203--3220},
publisher = {IEEE},
url        = {https://arxiv.org/abs/2311.10825}
}

@inproceedings{kesdogan1998stop,
  title={Stop-and-go-mixes providing probabilistic anonymity in an open system},
  author={Kesdogan, Dogan and Egner, Jan and B{\"u}schkes, Roland},
  booktitle={International Workshop on Information Hiding},
  pages={83--98},
  year={1998},
  organization={Springer}
}

@inproceedings{zantout2011i2p,
  title={I2P data communication system},
  author={Zantout, Bassam and Haraty, Ramzi and others},
  booktitle={Proceedings of ICN},
  pages={401--409},
  year={2011},
  organization={Citeseer}
}

@inproceedings{Mazieres1998Nymserver,
author    = {David Mazi{\`e}res and M. Frans Kaashoek},
title     = {The Design, Implementation and Operation of an Email Pseudonym Server},
booktitle = {Proceedings of the 5th ACM Conference on Computer and Communications Security (CCS)},
year      = {1998},
pages     = {27--36},
publisher = {ACM},
doi       = {10.1145/288090.288098}
}

@inproceedings{karakostas2024blockchain,
  title={Blockchain bribing attacks and the efficacy of counterincentives},
  author={Karakostas, Dimitris and Kiayias, Aggelos and Zacharias, Thomas},
  booktitle={Proceedings of the 2024 on ACM SIGSAC Conference on Computer and Communications Security},
  pages={1031--1045},
  year={2024}
}

@article{chaum1981untraceable,
  title={Untraceable electronic mail, return addresses, and digital pseudonyms},
  author={Chaum, David L},
  journal={Communications of the ACM},
  volume={24},
  number={2},
  pages={84--90},
  year={1981},
  publisher={ACM New York, NY, USA}
}

@inproceedings{diaz2004taxonomy,
  title={Taxonomy of mixes and dummy traffic},
  author={Diaz, Claudia and Preneel, Bart},
  booktitle={Information Security Management, Education and Privacy: IFIP 18th World Computer Congress TC11 19th International Information Security Workshops 22--27 August 2004 Toulouse, France},
  pages={217--232},
  year={2004},
  publisher={Springer},
  address={Boston, MA, USA},
  organization={Springer}
}

@inproceedings{piotrowska2017loopix,
  title={The loopix anonymity system},
  author={Piotrowska, Ania M and Hayes, Jamie and Elahi, Tariq and Meiser, Sebastian and Danezis, George},
  booktitle={26th USENIX Security Symposium (USENIX Security 17)},
  pages={1199--1216},
  year={2017},
  publisher={USENIX Association},
  address={Vancouver, BC, Canada}
}

@article{oldenburg2024mixmatch,
  title={Mixmatch: Flow matching for mixnet traffic},
  author={Oldenburg, Lennart and Juarez, Marc and R{\'u}a, Enrique Argones and Diaz, Claudia},
  journal={Proceedings on Privacy Enhancing Technologies},
  year={2024}
}

@inproceedings{syverson2004tor,
  title={Tor: The secondgeneration onion router},
  author={Syverson, Paul and Dingledine, Roger and Mathewson, Nick},
  booktitle={Usenix Security},
  volume={10},
  year={2004},
  organization={USENIX Association Berkeley, CA}
}

@inproceedings{danezis2003mixminion,
  title={Mixminion: Design of a type III anonymous remailer protocol},
  author={Danezis, George and Dingledine, Roger and Mathewson, Nick},
  booktitle={2003 Symposium on Security and Privacy, 2003.},
  pages={2--15},
  year={2003},
  organization={IEEE}
}

@article{shirvani2024past,
  title={The past, present, and future of the internet: A statistical, technical, and functional comparison of wired/wireless fixed/mobile internet},
  author={Shirvani Moghaddam, Shahriar},
  journal={Electronics},
  volume={13},
  number={10},
  pages={1986},
  year={2024},
  publisher={MDPI}
}

@article{diaz2022reward,
  title={Reward sharing for mixnets},
  author={Diaz, Claudia and Halpin, Harry and Kiayias, Aggelos},
  year={2022},
  publisher={Metagov}
}

@article{ccalicskan2015technical,
  title={Technical and legal overview of the tor anonymity network},
  author={{\c{C}}al{\i}{\c{s}}kan, Emin and Min{\'a}rik, Tom{\'a}{\v{s}} and Osula, Anna-Maria},
  journal={NATO Cooperative Cyber Defence Centre of Excellence. Available: https://ccdcoe. org/sites/default/files/multimedia/pdf/TOR\_Anonymity\_Network. pdf (January 4, 2016)},
  year={2015}
}

@article{jia2025kerblam,
  title={Kerblam—Anonymous messaging system protecting both senders and recipients},
  author={Jia, Yanxue and Das, Debajyoti and Zhang, Wenhao and Kate, Aniket},
  journal={Cryptology ePrint Archive},
  year={2025}
}

@inproceedings{hoang2018empirical,
  title={An empirical study of the i2p anonymity network and its censorship resistance},
  author={Hoang, Nguyen Phong and Kintis, Panagiotis and Antonakakis, Manos and Polychronakis, Michalis},
  booktitle={Proceedings of the internet measurement conference 2018},
  pages={379--392},
  year={2018}
}

@article{troncoso2017systematizing,
  title={Systematizing decentralization and privacy: Lessons from 15 years of research and deployments},
  author={Troncoso, Carmela and Isaakidis, Marios and Danezis, George and Halpin, Harry},
  journal={arXiv preprint arXiv:1704.08065},
  year={2017}
}

@inproceedings{egger2013practical,
  title={Practical attacks against the I2P network},
  author={Egger, Christoph and Schlumberger, Johannes and Kruegel, Christopher and Vigna, Giovanni},
  booktitle={International workshop on recent advances in intrusion detection},
  pages={432--451},
  year={2013},
  organization={Springer}
}

@inproceedings{herrmann2011privacy,
  title={Privacy-implications of performance-based peer selection by onion-routers: a real-world case study using I2P},
  author={Herrmann, Michael and Grothoff, Christian},
  booktitle={International Symposium on Privacy Enhancing Technologies Symposium},
  pages={155--174},
  year={2011},
  organization={Springer}
}

@inproceedings{jeong2016longitudinal,
  title={A Longitudinal Analysis of. i2p Leakage in the Public DNS Infrastructure.},
  author={Jeong, Seong Hoon and Kang, Ah Reum and Kim, Joongheon and Kim, Huy Kang and Mohaisen, Aziz},
  booktitle={SIGCOMM},
  volume={16},
  pages={557--558},
  year={2016}
}

@book{pathak2024web,
  title={Web Automation Testing Using Playwright: End-to-end, API, accessibility, and visual testing using Playwright (English Edition)},
  author={Pathak, Kailash},
  year={2024},
  publisher={Bpb Publications}
}

@inproceedings{platzer2020critical,
  title={Critical traffic analysis on the tor network},
  author={Platzer, Florian and Sch{\"a}fer, Marcel and Steinebach, Martin},
  booktitle={Proceedings of the 15th International Conference on Availability, Reliability and Security},
  pages={1--10},
  year={2020}
}

@article{constantinides2026ve,
  title={I've Seen This IP: A Practical Intersection Attack Against Tor Introduction Circuits and Hidden Services},
  author={Constantinides, Nicolas},
  journal={arXiv e-prints},
  pages={arXiv--2602},
  year={2026}
}

@inproceedings{johnson2013users,
  title={Users get routed: Traffic correlation on Tor by realistic adversaries},
  author={Johnson, Aaron and Wacek, Chris and Jansen, Rob and Sherr, Micah and Syverson, Paul},
  booktitle={Proceedings of the 2013 ACM SIGSAC conference on Computer \& communications security},
  pages={337--348},
  year={2013}
}

@inproceedings{kwon2020xrd,
  title={XRD: scalable messaging system with cryptographic privacy},
  author={Kwon, Albert and Lu, David and Devadas, Srinivas},
  booktitle={Proceedings of the 17th Usenix Conference on Networked Systems Design and Implementation},
  pages={759--776},
  year={2020}
}

@inproceedings{van2015vuvuzela,
  title={Vuvuzela: Scalable private messaging resistant to traffic analysis},
  author={Van Den Hooff, Jelle and Lazar, David and Zaharia, Matei and Zeldovich, Nickolai},
  booktitle={Proceedings of the 25th Symposium on Operating Systems Principles},
  pages={137--152},
  year={2015}
}

@inproceedings{mahdi2024larmix,
  title={LARMix: Latency-Aware Routing in Mix Networks},
  author={Rahimi, Mahdi and Sharma, Piyush Kumar and Diaz, Claudia},
  booktitle={The Network and Distributed System Security Symposium},
  year={2024},
  organization={Internet Society}
}

@inproceedings{mahdi2025lamp,
  title={LAMP: Lightweight Approaches for Latency Minimization in Mixnets with Practical Deployment Considerations},
  author={Rahimi, Mahdi and Sharma, Piyush Kumar and Diaz, Claudia},
  booktitle={The Network and Distributed System Security Symposium},
  year={2025},
  organization={Internet Society}
}

@inproceedings{mahdi2026OptiMix,
  title={{OptiMix}: Scalable and Distributed Approaches for Latency Optimization in Modern Mixnets},
  author={Rahimi, Mahdi},
  booktitle={The Network and Distributed System Security Symposium},
  year={2026},
  organization={Internet Society}
}

@inproceedings{rahimi2025parsan,
  title={{PARSAN-Mix}: Packet-Aware Routing and Shuffling with Additional Noise for Latency Optimization in Mix Networks},
  author={Rahimi, Mahdi},
  booktitle={International Conference on Applied Cryptography and Network Security},
  pages={159--188},
  year={2025},
  organization={Springer}
}

@inproceedings{tyagi2017stadium,
  title={Stadium: A distributed metadata-private messaging system},
  author={Tyagi, Nirvan and Gilad, Yossi and Leung, Derek and Zaharia, Matei and Zeldovich, Nickolai},
  booktitle={Proceedings of the 26th Symposium on Operating Systems Principles},
  pages={423--440},
  year={2017}
}

\appendix
\color{black}
\section{Sphinx and SURB Construction Details}
\label{app:sphinx}

This appendix specifies the Sphinx packet construction used in Nym.
Section~\ref{sec:sphinx-surb} covers SURB replies and acknowledgments in the main text; here we give notation, header generation, and the last-mix-hop key-withholding construction (Appendix~\ref{app:last-hop-keys}) needed for the attacks and defenses.

\subsection{Notation and Packet Format}
\label{app:sphinx-notation}

We adopt some of the notation of Danezis and Goldberg~\cite{Danezis2009Sphinx}.
Let $\kappa$ denote the security parameter and let $G$ be a prime-order cyclic group of order $q$ with generator $g$, where the Decisional Diffie--Hellman assumption holds.
We write $G^*$ for the non-identity elements of $G$.

Each mix node $n_i \in \mathcal{N}$ has private key $x_{n_i}\in\mathbb{Z}_q^*$ and public key $y_{n_i}=g^{x_{n_i}}$, with $|y_{n_i}|=2\kappa$.
A next-hop routing address $n_{i+1}$ encodes network-layer forwarding information (address, port, padding) and is distinct from $y_{n_{i+1}}$; we assume $|n_{i+1}|=2\kappa$.
A destination identifier $\Delta$ satisfies $|\Delta|=2\kappa$.

Let $r$ be the maximum route length and $\nu\le r$ the actual length of ordered route $(n_0,\ldots,n_{\nu-1})$.
The construction uses key-derivation functions $h_\mu,h_\rho,h_\sigma:G^*\rightarrow\{0,1\}^{\kappa}$, payload-key hash $h_{\mathsf{k}}:\{0,1\}^{\kappa}\rightarrow\{0,1\}^{*}$, replay hash $h_\tau:G^*\rightarrow\{0,1\}^{2\kappa}$, and blinding function $h_b:G^*\times G^*\rightarrow\mathbb{Z}_q^*$.
Sphinx employs MAC $\mu$ for integrity, stream cipher $\rho$ for routing-block masking, and keyed permutation family $\pi$ for layered payload encryption.
We treat $\rho$ as a keyed pseudorandom generator $\rho:\{0,1\}^{\kappa}\rightarrow\{0,1\}^{(r+1)\eta}$.
For hop $i$ with shared secret $s_i\in G^*$, define stream $\rho_i=\rho(h_\rho(s_i))$; the payload-key seed is $\sigma_i=h_\sigma(s_i)$ and the payload key is $k_i^\pi=h_{\mathsf{k}}(\sigma_i)$ (Section~\ref{par:sphinx-payload-keys}).
With Nym's parameters ($\kappa=128$ bits, $r=5$), the on-wire Sphinx header size is $|M|=348$\,bytes.

The payload permutation is
\[
\pi:\{0,1\}^{\kappa}\times\{0,1\}^{\ell}\rightarrow\{0,1\}^{\ell},
\]
with Sphinx padding\footnote{Classic Sphinx prefixes a $\kappa$-bit zero string before $P$, so Nym's wire format still budgets $\kappa{+}1$ bytes ($17$\,B for $\kappa{=}128$ bits) of payload framing beyond the application plaintext. We omit that prefix here; our $\mathrm{Pad}_{\mathsf{POTP}}$ construction (Section~\ref{sec:replier-anon}) likewise uses only the $\texttt{0x01}$ separator, optionally followed by an embedded-header suffix.}
\[
\mathrm{Pad}_\pi(P,\ell)=P\Vert\texttt{0x01}\Vert 0^{\ell-|P|-1}.
\]
Let $\mathsf{Alg}=(\mathrm{Enc},\mathrm{Dec})$ denote a symmetric encryption scheme; we write $\mathrm{Enc}^{\mathsf{Alg}}_k$ and $\mathrm{Dec}^{\mathsf{Alg}}_k$ for encryption and decryption under key $k$.

Per-hop forward metadata is $m_i=f_i\Vert v_i\Vert d_i$ with $|m_i|=c$; on the final hop, $d_i$ is omitted and $\bar{m}=f_{\nu-1}\Vert v_{\nu-1}$ with $|\bar{m}|=c_f<c$.
The routing flag $f_i$ distinguishes hop kinds, including $\textsf{FORWARD}$, $\textsf{FINAL}$, $\textsf{SWAP\_HEADER}$ (Section~\ref{sec:replier-anon}), and $\textsf{EMBEDDED\_KEYS}$ (Appendix~\ref{app:last-hop-keys}).
The routing-block step size is $\eta:=3\kappa+c$.
We write $\operatorname{head}_i(x)$ and $\operatorname{tail}_i(x)$ for prefix and suffix of length~$i$, $\Vert$ for concatenation, $\epsilon$ for the empty string, and $|x|$ for bit-string length.
A Sphinx header at hop $i$ is $M_i=(\alpha_i,\beta_i,\gamma_i)$, where $\alpha_i\in G$ is an ephemeral Diffie--Hellman element, $\beta_i$ is an encrypted routing block, and $\gamma_i$ is an integrity tag over $\beta_i$.
A complete Sphinx packet is $(M_0,\delta_0)$, where $\delta_0$ is the outermost payload ciphertext.

\subsection{Header and Routing-Block Construction}
\label{app:sphinx-header}

To construct a header, the sender takes destination $\Delta$, packet identifier $I$, and ordered route with public keys $((n_0,y_{n_0}),\ldots,(n_{\nu-1},y_{n_{\nu-1}}))$.
It samples $x\xleftarrow{\$}\mathbb{Z}_q^*$ and sets
\[
\alpha_0=g^x,
\qquad
s_0=y_{n_0}^{x},
\qquad
b_0=h_b(\alpha_0,s_0).
\]
For each $i>0$,
\[
\alpha_i=g^{x\prod_{j=0}^{i-1}b_j},
\qquad
s_i=y_{n_i}^{x\prod_{j=0}^{i-1}b_j},
\qquad
b_i=h_b(\alpha_i,s_i).
\]

\begin{figure}[!tb]
    \centering
    \adjustbox{max width=\columnwidth,max height=0.30\textheight,center}{%
      % Sphinx filler recursion for Nym's parameters (nu=4, r=5).
% Layout + routing parameters — adjust only this header block.

\newdimen\cell
\cell=0.40cm

\def\rowGap{0.62\cell}        % uniform gap: ρ→φ, concat→ρ within a stage
\def\xferGap{0.95\cell}        % φ output → φ|0 concat row
\def\dimEtaLift{0.28\cell}
\def\dimSixLift{0.18\cell}
\def\xorOff{0.80\cell}        % XOR west anchor offset east of colR
\def\xorLane{0.30\cell}        % fixed vertical bend rail east of 0 / concat inputs
\def\xorPhiHook{0.18\cell}     % visible vertical hook above φ.north (horizontal sits here)
\def\opLabelSep{3pt}
\def\boxLineWidth{0.45pt}
\def\markLineWidth{0.45pt}
\def\stealthLen{1.2mm}

\def\totalEta{6}

\newcommand{\rhoBandW}[1]{\dimexpr \numexpr#1+1\relax\cell\relax}
\newcommand{\shadeW}[1]{%
  \dimexpr \numexpr\totalEta-\numexpr#1+1\relax\relax\cell}

% --- orthogonal routing (identical on stages s0, s1, s2) ---
\newcommand{\placeXor}[2]{%
  \node[xor, anchor=west] (xor#1) at (xorRail |- #2) {};
  \node[right=\opLabelSep of xor#1, font=\scriptsize, inner sep=1pt] {XOR};
}
\newcommand{\routeInToXor}[3]{%
  \draw[arr] (#2.east) -- (laneX |- #2.east)
            -- (laneX |- #3.north) -- (#3.north);
}
\newcommand{\routeRhoToXor}[3]{%
  \draw[arr] (#2.east) -- (#3.west);
}
\newcommand{\routeXorToPhi}[3]{%
  \coordinate (hookY-#1) at ([yshift=\xorPhiHook]#3.north);
  \draw[arr] (#2.south) -- (xorCX |- hookY-#1)
            -- (#3.north |- hookY-#1) -- (#3.north);
}
\newcommand{\routeXfer}[3]{%
  \coordinate (midY-#1) at ($(#2.south west)!0.5!(#3.north)$);
  \draw[arr] (#2.south west) -- (#2.south west |- midY-#1)
            -- (#3.north |- midY-#1) -- (#3.north);
}

\begin{tikzpicture}[
  font=\footnotesize,
  smallbox/.style={
    draw,
    line width=\boxLineWidth,
    fill=gray!12,
    minimum width=\cell,
    minimum height=\cell,
    inner sep=0pt,
    outer sep=0pt,
    align=center,
  },
  pairbox/.style={
    draw,
    line width=\boxLineWidth,
    fill=gray!12,
    minimum height=\cell,
    inner sep=0pt,
    outer sep=0pt,
    align=center,
  },
  shadowbar/.style={
    draw,
    line width=\boxLineWidth,
    fill=gray!28,
    minimum height=\cell,
  },
  xor/.style={
    circle,
    draw,
    line width=\boxLineWidth,
    fill=white,
    minimum size=0.50cm,
    inner sep=0pt,
    font=\scriptsize,
  },
  arr/.style={-{Stealth[length=\stealthLen]}, line width=\boxLineWidth},
]

% Global alignment rails (shared by every stage)
\coordinate (colR) at (3.4*\cell, 0);
\coordinate (xorRail) at ([xshift=\xorOff]colR);
\coordinate (xorCX) at ([xshift=0.25cm]xorRail);
\coordinate (laneX) at ([xshift=\xorLane]colR);

% ===== Stage 0: 0 XOR ρ₀ → φ₁ =====
\node[smallbox, anchor=north east] (zero) at (colR) {0};

\coordinate (zL) at ([yshift=\dimEtaLift]zero.north west);
\coordinate (zR) at ([yshift=\dimEtaLift]zero.north east);
\draw[line width=\markLineWidth] (zL) -- (zR)
  node[midway, above=1pt, font=\scriptsize, inner sep=1pt] {$\eta$};

\node[smallbox, anchor=north east] (rho0)
  at ([yshift=-\rowGap]zero.south -| colR) {$\rho_0$};
\node[shadowbar, minimum width=\shadeW{0}, anchor=north east, left=0pt of rho0] (shade0) {};

\coordinate (wL0) at ([yshift=\dimSixLift]shade0.north west);
\coordinate (wR0) at ([yshift=\dimSixLift]rho0.north east);
\draw[line width=\markLineWidth] (wL0) -- (wR0)
  node[midway, above=1pt, font=\scriptsize, inner sep=1pt] {$6\eta$};

\placeXor{0}{rho0}

\node[smallbox, anchor=north east] (phi1)
  at ([yshift=-\rowGap]rho0.south -| colR) {$\phi_1$};

\routeInToXor{s0}{zero}{xor0}
\routeRhoToXor{s0}{rho0}{xor0}
\routeXorToPhi{s0}{xor0}{phi1}

% ===== Stage 1: φ₁|0 XOR ρ₁ → φ₂ =====
\node[smallbox, anchor=north east] (z1)
  at ([yshift=-\xferGap]phi1.south -| colR) {0};
\node[smallbox, anchor=east, left=0pt of z1] (phi1in) {$\phi_1$};

\routeXfer{x1}{phi1}{phi1in}

\node[pairbox, minimum width=\rhoBandW{1}, anchor=north east]
  (rho1) at ([yshift=-\rowGap]phi1in.south -| colR) {$\rho_1$};
\node[shadowbar, minimum width=\shadeW{1}, anchor=north east, left=0pt of rho1] (shade1) {};

\placeXor{1}{rho1}

\node[pairbox, minimum width=\rhoBandW{1}, anchor=north west]
  (phi2) at ([yshift=-\rowGap]rho1.south west) {$\phi_2$};

\routeInToXor{s1}{z1}{xor1}
\routeRhoToXor{s1}{rho1}{xor1}
\routeXorToPhi{s1}{xor1}{phi2}

% ===== Stage 2: φ₂|0 XOR ρ₂ → φ₃ =====
\node[smallbox, anchor=north east] (z2)
  at ([yshift=-\xferGap]phi2.south -| colR) {0};
\node[pairbox, minimum width=\rhoBandW{1}, anchor=east, left=0pt of z2] (phi2in) {$\phi_2$};

\routeXfer{x2}{phi2}{phi2in}

\node[pairbox, minimum width=\rhoBandW{2}, anchor=north east]
  (rho2) at ([yshift=-\rowGap]phi2in.south -| colR) {$\rho_2$};
\node[shadowbar, minimum width=\shadeW{2}, anchor=north east, left=0pt of rho2] (shade2) {};

\placeXor{2}{rho2}

\node[pairbox, minimum width=\rhoBandW{2}, anchor=north east]
  (phi3) at ([yshift=-\rowGap]rho2.south -| colR) {$\phi_3$};

\routeInToXor{s2}{z2}{xor2}
\routeRhoToXor{s2}{rho2}{xor2}
\routeXorToPhi{s2}{xor2}{phi3}

\end{tikzpicture}%
    }
    \caption{Recursive construction of Sphinx filler strings $\phi_i$ for $\nu=4$ and $r=5$.
    Here $\rho_i$ denotes the final $(i{+}1)\eta$ bits of $\rho(h_\rho(s_i))$.
    Filler strings pad routing blocks to fixed length $r\eta$ regardless of actual route length~$\nu$. Diagram drawn with AI assistance to the authors' specification.}
    \label{fig:sphinx-filler}
\end{figure}
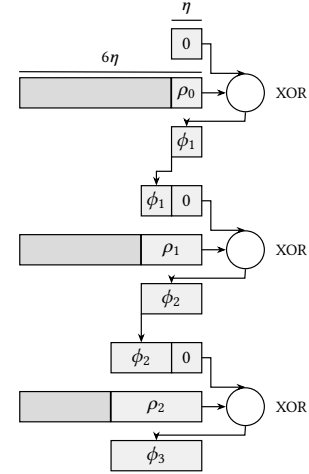

\begin{figure}[!tb]
    \centering
    \adjustbox{width=\columnwidth,max height=0.52\textheight,center}{%
      \input{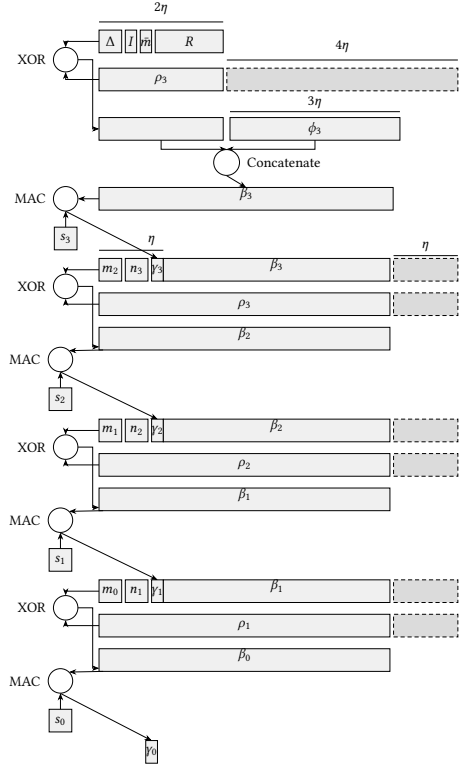}%
    }
    \caption{Recursive construction of Sphinx routing blocks for $\nu=4$, $r=5$, and $|\Delta|=2\kappa$.
    Each row shows the XOR masking of routing metadata with $\operatorname{head}(\rho_i)$; integrity tags $\gamma_i$ authenticate the resulting $\beta_i$. Diagram drawn with AI assistance to the authors' specification.}
    \label{fig:sphinx-header}
\end{figure}

\paragraph{Filler strings.}
Starting from $\phi_0=\epsilon$, the sender computes
\[
\phi_i=(\phi_{i-1}\Vert 0_{\eta})\oplus\operatorname{tail}_{i\eta}(\rho_{i-1}),
\qquad
0<i<\nu.
\]
Figure~\ref{fig:sphinx-filler} illustrates this recursion; shaded regions in Figure~\ref{fig:sphinx-header} mark portions of $\rho(h_\rho(s_i))$ not consumed by XOR masking.

\paragraph{Routing blocks.}
Following Loopix~\cite{piotrowska2017loopix}, for each forward hop the sender samples delay $d_i\sim\mathrm{Exp}(\mu)$ with $\mathbb{E}[d_i]=\mu$, sets $m_i=f_i\Vert v_i\Vert d_i$, and on the final hop uses $\bar{m}=f_{\nu-1}\Vert v_{\nu-1}$.
Routing blocks $M_{\nu-1},\ldots,M_0$ are built recursively.
The final block is
\begin{align}
R_{\nu-1}&\xleftarrow{\$}\{0,1\}^{(r-\nu+1)\eta-c_f-3\kappa}, \notag\\
\beta_{\nu-1}
&=
\Big(
(\bar{m}\Vert\Delta\Vert I\Vert R_{\nu-1})
\oplus
\operatorname{head}_{(r-\nu+1)\eta}(\rho_{\nu-1})
\Big)
\Vert
\phi_{\nu-1},
\end{align}
with $|R_{\nu-1}|$ chosen so that $|\beta_{\nu-1}|=r\eta$.
For $0\le i<\nu-1$,
\begin{align}
\beta_i
&=
\Big(
m_i\Vert n_{i+1}\Vert\gamma_{i+1}\Vert\operatorname{head}_{(r-1)\eta}(\beta_{i+1})
\Big)
\oplus
\operatorname{head}_{r\eta}(\rho_i), \\
\gamma_i&=\mu(h_\mu(s_i),\beta_i).
\end{align}
The resulting header is $M_0=(\alpha_0,\beta_0,\gamma_0)$.

\paragraph{Payload encryption.}
\label{par:sphinx-payload-keys}
Payload keys are derived from the per-hop Diffie--Hellman secrets by
\begin{align}
\label{eq:payload-seed}
\sigma_i&=h_\sigma(s_i)\in\{0,1\}^\kappa,\\
\label{eq:payload-key}
k_i^\pi&=h_{\mathsf{k}}(\sigma_i),\qquad i=0,\ldots,\nu-1,
\end{align}
where $h_\sigma$ hashes the shared secret to a seed and $h_{\mathsf{k}}$ hashes the seed to a payload key.
For plaintext $P$, the sender forms $P'=\mathrm{Pad}_\pi(P,\ell)$ and computes
\[
\delta_0=\pi_{k_0^\pi}\circ\cdots\circ\pi_{k_{\nu-1}^\pi}(P').
\]
The packet $(M_0,\delta_0)$ is transmitted to the first mix hop $n_0$.
Each mix node removes one layer using the key $k_i^\pi$ re-derived locally from $s_i$ (Section~\ref{sec:sphinx-surb}).
When a SURB is shipped to a replier, the seeds $\sigma_0,\ldots,\sigma_{\nu-1}$ are transmitted rather than the derived keys; the replier applies~\eqref{eq:payload-key} before encrypting (Section~\ref{sec:sphinx-surb}).

\subsection{Pudding Deterministic SURB Generation}
\label{app:pudding-surb}

Pudding~\cite{Kocaogullar2024Pudding} adapts the SURB construction of Section~\ref{sec:sphinx-surb} by replacing every source of randomness with a pseudorandom generator seeded from $\mathrm{KDF}(\mathsf{nonce}\,\|\,\mathit{ID}\,\|\,k)$.
The queried identifier~$\mathit{ID}$, the nonce~$\mathsf{nonce}$ carried in the lookup message, and a long-lived secret~$k$ provisioned among the discovery nodes are the only entropy inputs; $\mathrm{KDF}$ denotes a key derivation function.
The mixnet route $(n_0,\ldots,n_{\nu-1})$ is derived from the same KDF output rather than sampled independently.
Given destination reachability information~$\Delta$, the modified procedure therefore takes $(\mathit{ID},\mathsf{nonce},\Delta)$ as input and yields a standard SURB whose header, payload keys, and route are fixed by the seed.
All honest discovery nodes generate the same SURB for the same triple, keeping responses consistent across replicas.
Users do not know~$k$ and cannot execute this procedure locally; a malicious querier thus cannot precompute the reachability SURB returned for a registered identifier and infer membership from the reply path alone.

\subsection{Last-Hop Key Withholding via the Final Routing Block}
\label{app:last-hop-keys}

\begin{table}[!tb]
\centering
\caption{Comparison of standard and replier-anonymous SURBs.}
\label{tab:surb-comparison}
\setlength{\tabcolsep}{3pt}
\footnotesize
\begin{tabular}{@{}l>{\raggedright\arraybackslash}p{0.29\columnwidth}>{\raggedright\arraybackslash}p{0.29\columnwidth}@{}}
\toprule
\textbf{Aspect} & \textbf{Standard SURB} & \textbf{Replier-anonymous SURB} \\
\midrule
Payload encryption &
Layered pseudorandom permutation &
Pseudo one-time pad \\
SURB contents &
Mix-derived payload-key seeds provided to the replier &
Reply key only; mix-derived seeds recovered at the creator's gateway \\
First-mix-hop recognizability &
First mix hop observes the gateway that injected the reply &
First mix hop observes the swap node forwarding the reply header \\
\bottomrule
\end{tabular}
\end{table}

This appendix gives the modified Sphinx header construction used by the last-mix-hop defense of Section~\ref{sec:last-hop-fix}. The payload construction follows the POTP reply format of Section~\ref{sec:replier-anon} (Eqs.~\eqref{eq:potp}--\eqref{eq:pad-potp}). Nym uses $r=5$ routing blocks while ordinary reply paths contain $\nu=4$ Sphinx hops (three mix nodes and the recipient gateway), leaving one unused routing step of size $\eta$ in the final routing block.

\paragraph{Modified Sphinx construction.}
Only the final routing block differs from the construction of Section~\ref{app:sphinx-header}. The SURB creator derives the shared secrets $s_i$ and payload seeds $\sigma_i=h_\sigma(s_i)$ as in Eqs.~\eqref{eq:payload-seed}--\eqref{eq:payload-key}, and forms
\[
K_{\mathsf{emb}}
=
\sigma_0\Vert\sigma_1\Vert\cdots\Vert\sigma_{\nu-2},
\]
where
\[
|K_{\mathsf{emb}}|=(\nu-1)\kappa
\]
($48$\,B for $\nu=4$). The value $K_{\mathsf{emb}}$ replaces part of the random filler in the unused capacity of the final routing block, and the final-hop routing flag is set to $\textsf{EMBEDDED\_KEYS}$:
\begin{align}
\bar{m}&=f_{\nu-1}\Vert v_{\nu-1},
\qquad
f_{\nu-1}=\textsf{EMBEDDED\_KEYS}, \notag\\
R_{\nu-1}'&\xleftarrow{\$}
\{0,1\}^{(r-\nu+1)\eta-c_f-3\kappa-|K_{\mathsf{emb}}|}, \notag\\
\beta_{\nu-1}
&=
\Big(
(\bar{m}\Vert\Delta\Vert I\Vert K_{\mathsf{emb}}\Vert R_{\nu-1}')
\oplus
\operatorname{head}_{(r-\nu+1)\eta}(\rho_{\nu-1})
\Big)
\Vert
\phi_{\nu-1},
\end{align}
where $|R_{\nu-1}'|$ is chosen so that $|\beta_{\nu-1}|=r\eta$.

\begin{figure}[!t]
    \centering
    \adjustbox{width=\columnwidth,max height=0.52\textheight,center}{%
      \input{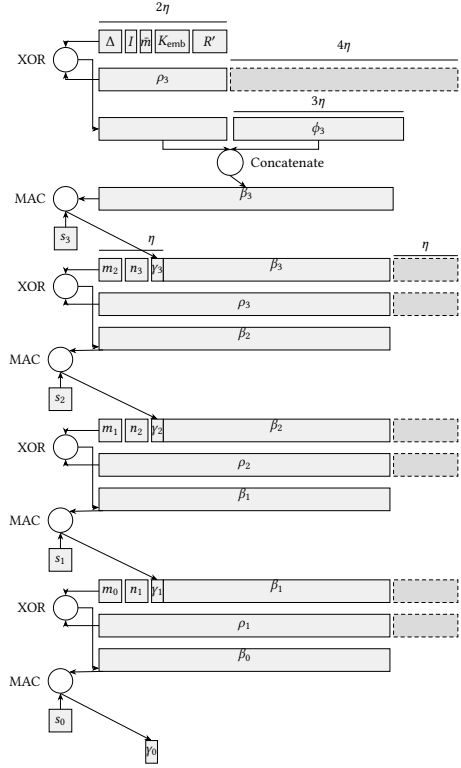}%
    }
    \caption{Recursive construction of Sphinx routing blocks with embedded prior-hop key seeds.
    Relative to Figure~\ref{fig:sphinx-header}, the final-hop filler $R$ is replaced by $K_{\mathsf{emb}}\Vert R'$ with $K_{\mathsf{emb}}=\sigma_0\Vert\sigma_1\Vert\sigma_2$ and $f_{\nu-1}=\textsf{EMBEDDED\_KEYS}$. Diagram drawn with AI assistance to the authors' specification.}
    \label{fig:sphinx-header-embedded}
\end{figure}

All preceding routing blocks are constructed exactly as in Section~\ref{app:sphinx-header}, using the $\textsf{FORWARD}$ flag, yielding the header $M_0=(\alpha_0,\beta_0,\gamma_0)$. Since $K_{\mathsf{emb}}$ appears only inside the onion-encrypted final routing block, a SURB holder that does not know the shared secrets $s_0,\ldots,s_{\nu-1}$ cannot recover the embedded payload seeds. Figure~\ref{fig:sphinx-header-embedded} illustrates the modification by replacing the final random filler $R$ of Figure~\ref{fig:sphinx-header} with $K_{\mathsf{emb}}\Vert R'$.
\paragraph{SURB creation.}
Per-hop POTP keys are derived as in~\eqref{eq:payload-key}:
\[
k_i^{\mathsf{POTP}}=h_{\mathsf{k}}(\sigma_i),
\qquad i=0,\ldots,\nu-1.
\]
Each mix node derives $\sigma_i=h_\sigma(s_i)$ locally from its shared secret, and the gateway similarly derives $\sigma_{\nu-1}=h_\sigma(s_{\nu-1})$; hence only $\sigma_0,\ldots,\sigma_{\nu-2}$ must be embedded in $K_{\mathsf{emb}}$. The creator samples a fresh reply key $\tilde{k}\xleftarrow{\$}\{0,1\}^{\kappa}$ and, instead of sending Nym's seed-augmented SURB
\[
\mathcal{S}
=
\bigl(
n_0,M_0,\sigma_0,\ldots,\sigma_{\nu-1},\tilde{k}
\bigr),
\]
sends only
\[
\mathcal{S}^{\star}
=
\bigl(
n_0,M_0,\tilde{k}
\bigr),
\]
where $M_0$ contains the embedded seeds in its modified final routing block. Thus the replier never receives the mix-derived seeds directly.

\paragraph{Reply encryption.}
For a reply fragment,
\[
p
=
\mathrm{SURB\text{-}ACK}
\Vert
H(\tilde{k})
\Vert
\mathrm{Enc}^{\mathsf{AES}}_{\tilde{k}}
(\mathsf{FH}\Vert f),
\]
The $\mathrm{SURB\text{-}ACK}$ and~$H(\tilde{k})$ remain outside the $\tilde{k}$ AES encrypted layer so that the gateway can recover and forward the ACK, and the client can match~$\tilde{k}$ by digest, without learning~$\tilde{k}$ at the gateway. The replier does not apply the mix-derived layers
\[
\mathrm{Enc}^{\mathsf{POTP}}_{k_0^{\mathsf{POTP}}}
\circ
\cdots
\circ
\mathrm{Enc}^{\mathsf{POTP}}_{k_{\nu-1}^{\mathsf{POTP}}},
\]
and sends $(M_0,\delta_0)$ to $n_0$.

Because the replier never learns $\sigma_0,\ldots,\sigma_{\nu-1}$, it cannot derive the mix-derived POTP keys or predict the payload ciphertext observed at the last mix hop, preventing last-mix-hop payload matching.

\paragraph{Mix and gateway processing.}
Hops $n_0,\ldots,n_{\nu-2}$ process the header as in Section~\ref{sec:sphinx-surb} and peel one POTP layer with $k_i^{\mathsf{POTP}}=h_{\mathsf{k}}(h_\sigma(s_i))$:
\[
\delta_{i+1}=\mathrm{Dec}^{\mathsf{POTP}}_{k_i^{\mathsf{POTP}}}(\delta_i).
\]
(The mix peels act on a payload that was not pre-wrapped under those keys by the replier; as in Danezis and Goldberg~\cite{Danezis2009Sphinx}, the destination undoes this mismatch using the recovered keys.)
At the final hop $n_{\nu-1}$ (the recipient gateway), Sphinx processing recovers $\bar{m}$ with $f_{\nu-1}=\textsf{EMBEDDED\_KEYS}$, together with $\Delta$, $I$, and $K_{\mathsf{emb}}=\sigma_0\Vert\cdots\Vert\sigma_{\nu-2}$.
The gateway derives $\sigma_{\nu-1}=h_\sigma(s_{\nu-1})$ (hence $k_{\nu-1}^{\mathsf{POTP}}=h_{\mathsf{k}}(\sigma_{\nu-1})$) from its own shared secret, derives keys from the embedded seeds by~\eqref{eq:payload-key},
\[
k_i^{\mathsf{POTP}}=h_{\mathsf{k}}(\sigma_i),\qquad i=0,\ldots,\nu-2,
\]
and peels its own POTP layer to obtain $\delta_{\nu-1}$.
Because $\mathrm{Enc}^{\mathsf{POTP}}=\mathrm{Dec}^{\mathsf{POTP}}$, it restores the padded plaintext \emph{except} for the $\tilde{k}$ body layer by
\[
\mathrm{Dec}^{\mathsf{POTP}}_{k_0^{\mathsf{POTP}}}
\circ
\cdots
\circ
\mathrm{Dec}^{\mathsf{POTP}}_{k_{\nu-1}^{\mathsf{POTP}}}
(\delta_\nu),
\]
forwards the clear $\mathrm{SURB\text{-}ACK}$ into the mixnet, and delivers $H(\tilde{k})\Vert\mathrm{Enc}^{\mathsf{AES}}_{\tilde{k}}(\mathsf{FH}\Vert f)$ to the client---including when the client is offline and the gateway must buffer the fragment.
The client alone holds~$\tilde{k}$: it matches $H(\tilde{k})$, and decrypts $\mathrm{Enc}^{\mathsf{AES}}_{\tilde{k}}(\mathsf{FH}\Vert f)$.
Thus gateway-issued acknowledgments are preserved without giving the gateway~$\tilde{k}$, while the replier never holds the mix-derived POTP keys required for the last-mix-hop attack of Section~\ref{sec:first-hop-surb-attack}.

\section{Implementation Details}
\label{app:implementation}

This appendix describes some implementation details of
Section~\ref{sec:replier-anon} and the protocol of Section~\ref{sec:hidden-service}.

\paragraph{Binaries.}
The prototype comprises three mixnet-connected binaries:
\begin{itemize}[nosep]
    \item \textbf{Client proxy} (\texttt{nym-sphinx-socks}): SOCKS5 termination, SURB retrieval, session state, and replier-anonymous sends.
    \item \textbf{Hidden service} (\texttt{hidden-service}): SURB publication, mixnet session termination, and proxying to a local origin server.
    \item \textbf{Repository node} (\texttt{repository-server}): signature verification, SURB inventory, and rendezvous blob storage.
\end{itemize}

\paragraph{Sphinx-layer changes.}
We replace the payload permutation $\pi$ with the
length-preserving POTP of Section~\ref{sec:replier-anon}, implemented as AES-128-CTR. The per-hop routing flag
$f_i$ then takes one of the values in Table~\ref{tab:routing-flags}, three of which NymHS
introduces. A high bit ($\mathtt{0x80}$) on any flag requests per-hop tracing, which we added for
debugging and to confirm that replier-anonymous SURBs route as intended.

An $\mathcal{S}^{\star}$ SURB carries the header and first-hop address alone. With
$\kappa=16$\,B and $r=5$ the header is $|M|=348$\,B, so at $\nu=4$ a replier receives
$396$\,B where the seed-carrying SURB deployed by Nym gives $460$\,B. The $64$\,B
difference is the mix-derived seeds whose disclosure to the replier enables the last
mix-hop attack of Section~\ref{sec:first-hop-surb-attack}. They travel in the final-hop
padding instead: a $\nu=4$ route leaves $68$\,B of padding, of which
$K_{\mathsf{emb}}$ occupies $49$\,B as a one-byte count followed by
$\sigma_0\Vert\sigma_1\Vert\sigma_2$.

\begin{table}[!tb]
\centering
\footnotesize
\setlength{\tabcolsep}{4pt}
\caption{Per-hop routing flags. The last three are introduced by NymHS.}
\label{tab:routing-flags}
\begin{tabular}{@{}llp{0.52\columnwidth}@{}}
\toprule
\textbf{Flag} & \textbf{Value} & \textbf{Effect at the node} \\
\midrule
$\textsf{FORWARD}$          & 1 & Peel one layer, delay, forward. \\
$\textsf{FINAL}$            & 2 & Recover $\Delta$ and the payload. \\
$\textsf{SWAP\_HEADER}$     & 3 & Extract $M_0$ from the payload tail, replace it with fresh randomness, forward under $M_0$. \\
$\textsf{EMBEDDED\_KEYS}$   & 4 & Parsed as $\textsf{FINAL}$, then recover $K_{\mathsf{emb}}$ and undo the mix-key mismatch. \\
$\textsf{ACK\_FINAL}$       & 6 & Parsed as $\textsf{FINAL}$, but only the leading $\mathrm{AckPacket}$ payload bytes are the acknowledgment. \\
\bottomrule
\end{tabular}
\end{table}

\begin{table}[!tb]
\centering
\caption{Authenticated hidden-service message format.}
\label{tab:hs-message}
\begin{tabular}{|l|>{\raggedright\arraybackslash}p{0.48\columnwidth}|}
\hline
\textbf{Field} & \textbf{Description} \\
\hline
$\mathsf{senderTag}$ & Ed25519 public key serving as a pseudonymous sender identifier. \\ \hline
$\mathsf{kind}$ & Message type: publication, retrieval, replenishment, session, or application data. \\ \hline
$\sigma$ & Ed25519 signature providing message authenticity and integrity. \\ \hline
$\mathsf{nonce}$ & Fresh nonce used to prevent replay attacks. \\ \hline
\end{tabular}
\end{table}

\begin{table}[!tb]
\centering
\footnotesize
\setlength{\tabcolsep}{2pt}
\renewcommand{\arraystretch}{1.15}
\caption{Prototype dispatch of $\mathsf{kind}$ values (paper names). CP\,=\,client proxy, HS\,=\,hidden service, RN\,=\,repository node.}
\label{tab:hs-kind-dispatch}
\begin{tabular}{@{}>{\raggedright\arraybackslash}p{0.28\columnwidth}>{\raggedright\arraybackslash}p{0.40\columnwidth}>{\raggedright\arraybackslash}p{0.22\columnwidth}@{}}
\toprule
\textbf{Message}  & \textbf{Purpose}  & \textbf{From\,$\to$\,To} \\
\midrule
$\mathsf{SignedSURBBundle}$  & Publish or return signed service SURBs keyed by $PK_S$.  & HS\,$\to$\,RN; RN\,$\to$\,CP \\
$\mathsf{SURBRequest}$  & Ask RN for $k$ fresh signed SURBs for $PK_S$.  & CP\,$\to$\,RN \\
$\mathsf{PublicationStatus}$  & Report publish/retrieve outcome.  & RN\,$\to$\,HS/CP \\
$\mathsf{SURBReplenish}$\newline$\mathsf{Request}$  & Ask HS to republish when inventory~$<k$.  & RN\,$\to$\,HS \\
$\mathsf{StoreRendezvous}$\newline$\mathsf{Blob}$  & Store encrypted replenishment SURB bundle.  & HS/CP\,$\to$\,RN \\
$\mathsf{PublicationAck}$  & Confirm blob stored under $\mathsf{rid}$.  & RN\,$\to$\,peer \\
$\mathsf{RendezvousLookup}$  & Fetch stored blob by $\mathsf{rid}$/nonce.  & CP/HS\,$\to$\,RN \\
$\mathsf{RendezvousBlob}$  & Return stored ciphertext to requester.  & RN\,$\to$\,req. \\
$\mathsf{Rendezvous}$\newline$\mathsf{Establish}$  & Start peer replenishment (DH + quota).  & req.\,$\to$\,peer \\
$\mathsf{RendezvousPointer}$  & Point requester to blob and reply budget.  & peer\,$\to$\,req. \\
$\mathsf{ApplicationRequest}$  & Authenticated application request.  & CP\,$\to$\,HS \\
$\mathsf{Application}$\newline$\mathsf{Response}$  & Signed application response on reply SURBs.  & HS\,$\to$\,CP \\
\bottomrule
\end{tabular}
\end{table}

\paragraph{Protocol realization.}
Table~\ref{tab:hs-kind-dispatch} lists the hidden-service messages the prototype
dispatches, one per value of the $\mathsf{kind}$ field of Table~\ref{tab:hs-message}.

The shared secret $\mathcal{U}$ of Section~\ref{sec:hs-publication} is a fresh $128$-bit
value drawn per repository at each publication and carried only in the service's own
upload; a repository serving SURBs to a client omits it. The repository stores $\mathcal{U}$ and SENDS it back when asking the service to republish, and the
service republishes only on a match. An envelope signature alone therefore does not
suffice to trigger a republication. A repository serves whatever inventory it holds once
at least $k$ SURBs are available, and requests replenishment when a request cannot be met
in full; outstanding replenishment requests expire after $60$\,s, so a service that
missed one is retried rather than stranded.

\section{Latency and Fragment Count}
\label{app:fragments}
Section~\ref{sec:eval-latency} attributes page-load latency to the number of Sphinx
fragments a page requires. We estimate that count by dividing the returned bytes by the
$\ell-445$ bytes of application data each fragment carries (Section~\ref{sec:eval-overhead}).
The estimate is a lower bound: resources are fragmented independently, so spreading the
same bytes across more resources can require additional fragments.

\begin{figure}[!tb]
\centering
\includegraphics[width=\columnwidth]{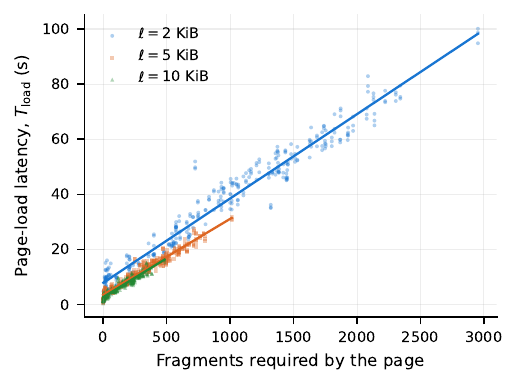}
\caption{Page-load latency versus estimated fragment count at $D=1$. Each point represents one measured load for each of the $118$ templates across three runs; lines show least-squares fits.}
\label{fig:latency-vs-fragments}
\end{figure}

Latency is approximately linear in the estimated count at every payload size
(Figure~\ref{fig:latency-vs-fragments}). Least-squares fits give per-fragment slopes of
$30.6$, $27.5$, and $28.2$\,ms at $\ell=2$, $5$, and $10$\,KiB, so a page of $F$
fragments loads in approximately $7.9+0.031F$ seconds at $\ell=2$\,KiB and
$2.7+0.028F$ seconds at $\ell=10$\,KiB. At $\ell=2$\,KiB the residual standard
deviation is $4.3$\,s, against $21.7$\,s when load time is predicted from the corpus
mean alone.

\begin{figure}[!tb]
\centering
\includegraphics[width=\columnwidth]{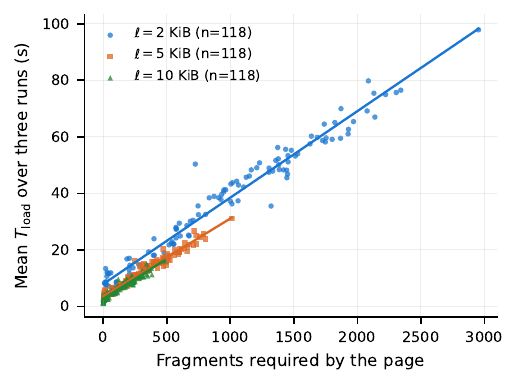}
\caption{Page-load latency versus estimated fragment count after averaging each template over its three runs, yielding $118$ points per payload size.}
\label{fig:latency-vs-fragments-avg}
\end{figure}

The residual is structural rather than noise: averaging each template over its three
runs (Figure~\ref{fig:latency-vs-fragments-avg}) reduces the residual standard
deviations only from $4.3$, $1.5$, and $1.0$\,s to $3.9$, $1.3$, and $0.9$\,s at
$\ell=2$, $5$, and $10$\,KiB. What the estimate does not capture---per-response
fragmentation granularity, and rendezvous replenishment rounds that install SURBs in
fixed blocks at the cost of a mixnet round trip independent of page size---accounts for
most of it.
%The fitted intercepts should not be interpreted as measured fixed costs. The fits yield intercepts of $7.9$,s at $\ell=2$,KiB and $2.7$,s at $\ell=10$,KiB, but the underlying relation is slightly curved, causing the linear fits to over-predict latency for the smallest pages, particularly at $\ell=2$,KiB. The two templates requiring fewer than five fragments, for example, load in $2.0$,s at $\ell=2$,KiB. Directly measured, the smallest pages take $2.0$,s at $\ell=2$,KiB and $2.3$,s at $\ell=10$,KiB, consistent with work that does not depend on~$\ell$.

%The fitted slopes identify the deployment-relevant effect: delivery cost is approximately per fragment rather than per byte. A fragment at $\ell=10$,KiB carries six times as much application data as one at $\ell=2$,KiB, yet their fitted per-fragment costs are similar ($28.2$ versus $30.6$,ms). This behavior follows from the Loopix design, which releases packets on a poisson rate irrespective of payload size. Increasing $\ell$ therefore amortizes a similar per-packet delivery cost over more application data, making the Sphinx payload size an important deployment parameter.

\section{Use Cases for Replier-Anonymous SURBs}
\label{app:usecases}

\subsection{Clearnet Browsing}
\label{app:clearnet-usecase}

As Section~\ref{sec:rw-nym-apps} notes, the exit gateway is the replier: the network
requester it runs embedded answers on the client's SURBs, and a replier creates the
acknowledgment carried in every response, chooses that acknowledgment's route, and may
operate its first hop~$n_0^{\mathsf{ACK}}$. Figure~\ref{fig:clearnet-usecase} shows what the rewrap of
Section~\ref{sec:replier-anon} changes: the exit keeps its route choice and the
acknowledgment its format, but an adversary operating~$n_0^{\mathsf{ACK}}$ receives the acknowledgment
header from the swap hop rather than from the client's entry gateway.

\begin{figure*}[t]
  \centering
  \begin{adjustbox}{max width=\textwidth}
    \input{figures/clearnet_usecase.tex}
  \end{adjustbox}

  \vspace{7pt}
  \begin{minipage}{0.97\textwidth}
  \footnotesize
  \begin{enumerate}[leftmargin=2.4em, itemsep=1.5pt, topsep=2pt,
                    label=\textbf{\arabic*.}]
    \item The client sends its request to the exit gateway, together with the SURBs on
      which it will be answered, over a route it selects itself.
    \item The exit gateway retrieves the page from the clearnet web server. This link is
      not anonymised; the server sees the gateway and not the client.
    \item The exit gateway returns the response on one of the client's SURBs. As the
      replier it also constructs the $\mathrm{SURB\text{-}ACK}$ carried inside that
      response, and therefore chooses the acknowledgment's route, including its first
      mix hop.
    \item The client's entry gateway extracts the $\mathrm{SURB\text{-}ACK}$. Instead of
      injecting it on the route its creator chose, the gateway rewraps it in an outer
      shell of depth $D=3$ whose hops the gateway selects itself.
    \item At the swap hop in L3 the outer header is consumed and the creator's inner
      header is installed, so the acknowledgment leaves the shell and continues on the
      route the exit gateway chose.
    \item The first hop $n_0^{\mathsf{ACK}}$ of that inner route may be operated by the exit gateway,
      which recognises the header it created. Its upstream neighbour is the layer-3
      mixnode of step~5 rather than the client's entry gateway, so the observation the
      attack depends on is unavailable.
  \end{enumerate}
  \end{minipage}
  \caption{Replier-anonymous SURBs in clearnet browsing. Each hop is outlined in the
  colour of the party that selected it. The acknowledgment is the only route whose two
  halves have different owners, and because neither owner observes the other's half,
  neither can recognise the packet as it enters the mixnet. We assume that a single fragment suffices for the request and for the response. Diagram drawn with AI assistance to the authors' specification.}
  \label{fig:clearnet-usecase}
\end{figure*}

The rewrap changes nothing else. The exit gateway keeps its route choice, the
acknowledgment keeps its format, and the client keeps the SURB pool it already supplies;
only the point at which the acknowledgment enters the mixnet moves out of the creator's
control.

\subsection{Nym-Address Applications}
\label{app:nymaddr-usecase}

Between two addressed parties these attacks disclose nothing the address had not already
given away. The case that matters is the one of Section~\ref{sec:rw-nym-apps}: a client
that withholds its own address and supplies the SURBs on which it will be answered, which
makes the client the creator and the addressed party the replier, inverting the
assignment of Section~\ref{app:clearnet-usecase} and leaving the client's entry gateway
as the only attachment point still hidden. Figure~\ref{fig:nymaddr-usecase} shows how the
rewrap keeps it hidden.

\begin{figure*}[t]
  \centering
  \begin{adjustbox}{max width=\textwidth}
    \input{figures/nymaddr_usecase.tex}
  \end{adjustbox}

  \vspace{7pt}
  \begin{minipage}{0.97\textwidth}
  \footnotesize
  \begin{enumerate}[leftmargin=2.4em, itemsep=1.5pt, topsep=2pt,
                    label=\textbf{\arabic*.}]
    \item The client sends its request over a route it selects itself, whose last hop is
      the gateway named by the address. That gateway holds the correspondence between
      the address and the party attached to it, and forwards the request to the holder.
      The client supplies $\mathcal{S}^{\star}$ SURBs in place of its own address, so
      its own gateway is never named.
    \item The holder answers on one of those SURBs. Because the client supplied
      $\mathcal{S}^{\star}$ SURBs, the holder cannot predict the ciphertext the reply
      presents at the last mix hop, which closes the companion attack of
      Section~\ref{sec:last-hop-fix}.
    \item As the replier, the holder also constructs the $\mathrm{SURB\text{-}ACK}$
      carried in that reply, and therefore chooses its route including the first mix
      hop~$n_0^{\mathsf{ACK}}$, which may be a mixnode it operates.
    \item The client's gateway extracts the acknowledgment and rewraps it in an outer
      shell of depth $D=3$ whose hops the gateway selects itself.
    \item At the swap hop in L3 the outer header is consumed and the holder's inner
      header installed, so the acknowledgment continues on the route its creator chose.
    \item $n_0^{\mathsf{ACK}}$ recognises the header it created, but its upstream
      neighbour is the layer-3 mixnode of step~5 rather than the client's gateway.
  \end{enumerate}
  \end{minipage}
  \caption{Replier-anonymous SURBs when an anonymous client contacts a Nym-address
  holder. Roles are reversed with respect to Figure~\ref{fig:clearnet-usecase}: here the
  client creates the SURBs and the addressed party replies on them. Both attacks a
  replier can mount against a creator land on the client's gateway;
  $\mathcal{S}^{\star}$ closes the last-mix-hop attack and the rewrap shown here closes
  the first-mix-hop attack on the acknowledgment. We assume that a single fragment suffices for the request and for the response. Diagram drawn with AI assistance to the authors' specification.}
  \label{fig:nymaddr-usecase}
\end{figure*}

\end{document}